\documentclass{sig-alternate-2013}

\newfont{\mycrnotice}{ptmr8t at 7pt}
\newfont{\myconfname}{ptmri8t at 7pt}
\let\crnotice\mycrnotice%
\let\confname\myconfname%

\permission{}

\conferenceinfo{COSN'15,}{November 2--3, 2015, Palo Alto, California, USA.}

\copyrightetc{}
\crdata{}

\usepackage[
  pass,% keep layout unchanged 
]{geometry}

\usepackage{amsmath,amssymb}
\usepackage{graphicx}
\usepackage{latexsym}
\usepackage{url}
\usepackage{multirow}

\usepackage[table]{xcolor}
\usepackage{color}
\usepackage{algorithm,algorithmic}
\usepackage{caption}
  \DeclareCaptionType{copyrightbox}
  \usepackage{subcaption}
\newcommand{\U}{\mathcal{U}}
\newcommand{\usr}{u}
\newcommand{\fri}[1]{{\it f}(#1)}
\newcommand{\com}{{\it c}}
\newcommand{\Com}[1]{{\it C}(#1)}

\newcommand{\coment}[1]{{\it coment}(#1)}

\newcommand{\Ci}[1]{{\it CI}(#1)}

\usepackage[colorlinks,
            linkcolor=black,
            anchorcolor=black,
            citecolor=blue,
			urlcolor=black,
            bookmarks=true
            ]{hyperref}
\begin{document}

\title{Location Prediction:\\ Communities Speak Louder than Friends}

\numberofauthors{3} 
\author{
\alignauthor
Jun Pang\\
      \affaddr{University of Luxembourg}\\
      \affaddr{FSTC \& SnT}\\
      \email{jun.pang@uni.lu}
\alignauthor
Yang Zhang\\
      \affaddr{University of Luxembourg}\\
      \affaddr{FSTC}\\
      \email{yang.zhang@uni.lu}
}
\maketitle

% ------------------------------------------------
\begin{abstract}
% ------------------------------------------------
Humans are social animals, they interact with different communities of friends to conduct different activities.
The literature shows that human mobility is constrained by their social relations.
In this paper, we investigate the social impact of a person's communities on his mobility,
instead of all friends from his online social networks.
This study can be particularly useful,
as certain social behaviors are influenced by specific communities but not all friends.
To achieve our goal, we first develop a measure
to characterize a person's social diversity,
which we term `community entropy'.
Through analysis of two real-life datasets,
we demonstrate that a person's mobility
is influenced only by a small fraction of his communities
and the influence depends on the social contexts of the communities.
We then exploit machine learning techniques to predict users' future movement 
based on their communities' information.
Extensive experiments demonstrate the prediction's effectiveness.
 
% ------------------------------------------------
\end{abstract}
\category{H.2.8}{Database Management}{Database Applications}[Data mining]
\terms{Algorithms, theory, experiments}
\keywords{Human mobility, social networks, network communities}

% ==================================================================
\section{Introduction}
\label{sec:intro}
% ==================================================================
Humans are social animals,
everyone is a part of the society and gets influences from it.
For example, our daily behaviors, 
such as what types of music we listen to, where we have lunch on weekdays
and what activities we conduct on weekends,
are largely dependent on our social relations.
Normally, we categorize our social relations into different groups,
i.e., social communities, using different criteria and considerations.
By definition,
\emph{a community is a social unit of any size that shares common values}.\footnote{\url{http://en.wikipedia.org/wiki/Community}}
Typical communities include family, close friends, colleagues, etc.
In daily life, humans are engaged in various social environments,
and they interact with different communities depending on the environments.
For our specific behaviors, social influences,
in most of cases, are not from \emph{all our friends} but from \emph{certain communities}.
For example, we listen to similar types of music as our close friends, but not as our parents;
we have lunch together with our colleagues on weekdays, but not with our college friends living in another city;
on weekends we spend more time with family, but not with our colleagues.

Location-based social network services (LBSNs) have been booming during the past five years.
Nowadays,  it is common for a user to attach his location 
when he publishes a photo or a status using his online social network account.
Moreover, users may just share their locations, 
called \emph{check-in}, to tell their friends where they are or to engage in social games as in Foursquare.
Since these large amount of location and social relation data become available, 
studying human mobility
and its connection with social relationships becomes quantitatively achievable (e.g.,~\cite{GHB07,CML11,CCLS11,SNLM11,GTL12,CPX13,CPX14}).
Understanding human mobility can lead to compelling applications 
including location recommendation~\cite{ZZXM09,ZZXY10,ZZMXM11,GTHL13,LX13}, urban planning~\cite{ZLH13}, 
immigration patterns~\cite{CMA05},~etc. 

Previous works, including~\cite{BSM10,CS11,CML11,SKB12}, show that 
human mobility is influenced by social factors.
However, there is one common shortcoming: 
they all treat friends of users equally.
%First, in online social networks some of a user's friends are not his real friends, 
%and the links may be created randomly or due to other malicious purposes~\cite{BE07}.
%These friends have little influence on users' behaviors.
%Second, even only considering real friends, 
Similar to other social behaviors, 
in most cases mobility is influenced by specific communities but not all friends.
For example, the aforementioned colleagues can influence the place a user goes for lunch 
but probably have nothing to do with his weekend plans.
Meanwhile, where a user visits on weekends largely depends on his friends or family,
but not his colleagues.
Therefore, the impact on a user's mobility should be considered from
the perspectives of communities instead of all friends.
In a broader view, community is arguably the most useful resolution to study social networks~\cite{YML14}.

\medskip
\noindent\textbf{Contributions.}
In this paper, we aim to study the impact from communities on a user's mobility 
and predict his locations based on his community information.

First, we partition each users' friends into communities 
and propose a notion namely community entropy to quantify a user's social diversity.
Second, we analyze communities' influences on users' mobility
and our main conclusions include:
(1) communities' influences on users' mobility are stronger than their friends';
(2) each user is only influenced by a small number of his communities;
and (3) such influence is typically constrained by temporal and spatial contexts.
Third, we predict users' locations using their community information.
Experimental results on two real-life datasets with millions of location data 
show that the community-based predictor achieves a strong performance.

\medskip
\noindent\textbf{Organization.} 
After the introduction, we present a few preliminaries
and our datasets in Section~\ref{sec:preli}.
Then we describe the community detection process and 
propose the notion of community entropy in Section~\ref{sec:comdet}.
The relationship between users and their communities on mobility is analyzed in Section~\ref{sec:commob}.
Based on our analysis, we propose a location predictor with features
linked to community information and present experimental results in Section~\ref{sec:locpre}.
We discuss related work in Section~\ref{sec:rel} and
conclude our paper with some future work in Section~\ref{sec:conclu}.

% ==================================================================
\section{Preliminaries}
\label{sec:preli}
% ==================================================================

We summarize the notations in Section~\ref{ssec:notations}
and describe the datasets that we use throughout the paper in Section~\ref{ssec:dataset}.

% ------------------------------------------------
\subsection{Notations}
\label{ssec:notations}
% ------------------------------------------------
All users are contained in the set $\U$ while a single user is denoted by $\usr$.
We use the set $\fri{\usr}$ to represent $\usr$'s friends.
A community of a user $\usr$ is a subset of his friends denoted by $\com$ and $\com\subseteq \fri{\usr}$.
Meanwhile, $\Com{\usr}$ represents all the communities of $\usr$, i.e., $\Com{\usr}$ is a set of sets of $\usr$'s communities.
Every friend of a user is assigned into one of the user's communities,
the union of all his communities is the set of all his friends.
In this work, we only consider non-overlapping communities,
namely $\com\cap\com' =\emptyset$ for $\com, \com' \in \Com{\usr}$.
However, this assumption is not crucial to our approach
and our results can be extended for overlapping communities as well. 

A check-in of $\usr$ is denoted by a tuple $\langle \usr, t, \ell \rangle$,
where $t$ represents the time 
and $\ell$ is the location that corresponds to a pair of latitude and longitude.
We use $\Ci{\usr}$ to represent all the check-ins of $\usr$.
Without ambiguity, we use location and check-in interchangeably in the following discussion.

\begin{figure*}[!t]
\begin{minipage}[b]{0.6\columnwidth}
  \centering
  \includegraphics[width=\columnwidth]{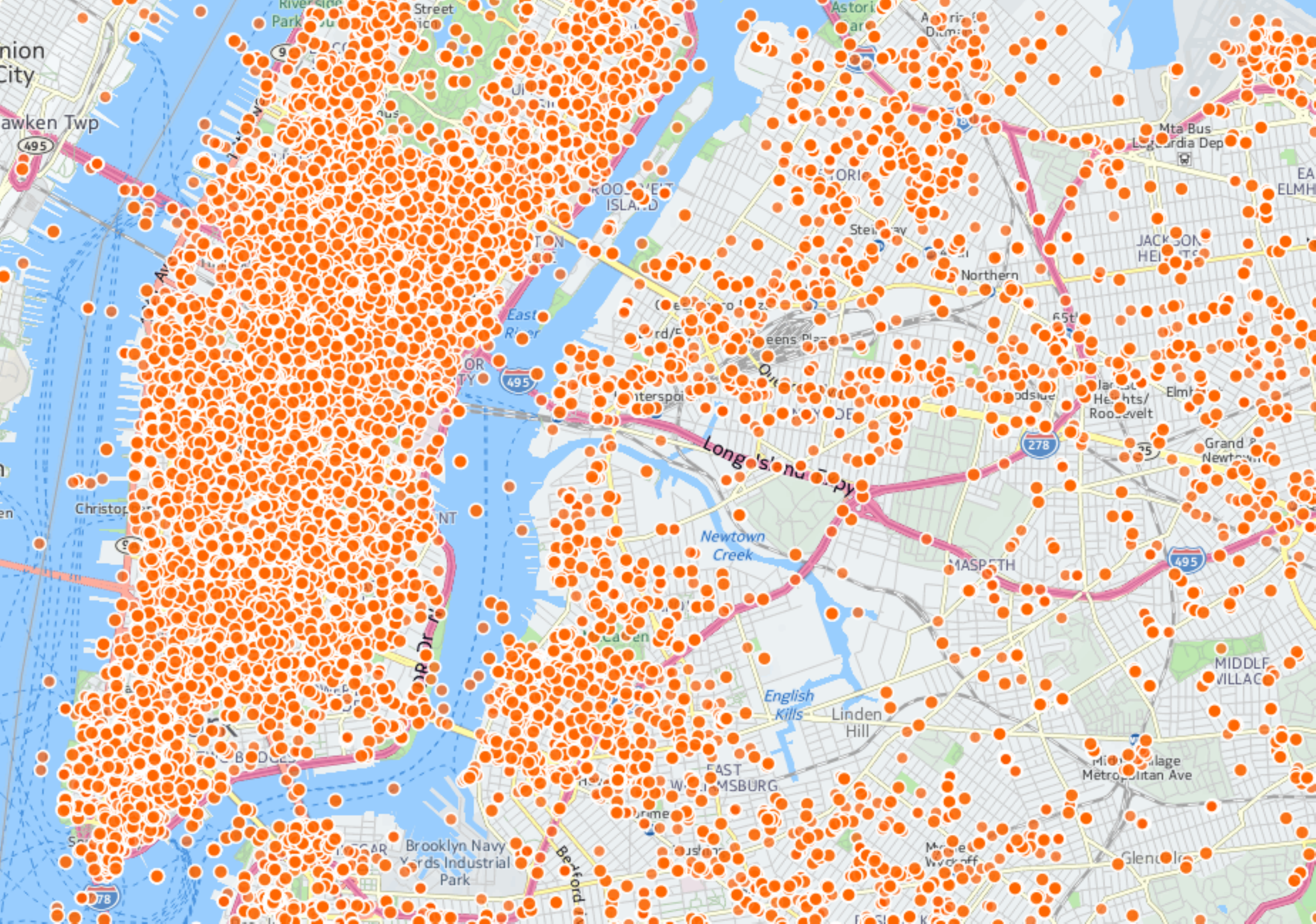}
  \caption{Check-ins in New York.}{\label{fig:cisample}}
\end{minipage}
\hspace{10mm}
\begin{minipage}[b]{0.6\columnwidth}
\centering
\def\arraystretch{1}
\setlength\tabcolsep{0.7mm}
\begin{tabular}{| c || c | c || c | c |}
  \hline
  & NY (G) & SF (G) & NY (T) & SF (T) \\
  \hline
  \hline
  \# of users & 7,786  & 6,617 & 207,805& 113,383\\
%   \hline
%   Avg.\# of friends & 21.9 &  25.2 & &\\
  \hline
  \# of check-ins & 176,324 & 177,357 & 2,325,907& 2,163,959\\
  \hline
  Avg.\# of check-ins & 21.6 &  26.8 & 11.2 &19.1\\
  \hline
  \# of active users & 175  & 236 & 1,636& 1,626\\
  \hline
  Avg.\# of friends (active user) & 79.4 & 69.7 & 376.9 & 289.0\\
  \hline
  \end{tabular}
\captionof{table}{Summary of the datasets. \label{tab:datasets}}
\end{minipage}
\end{figure*}

% ------------------------------------------------
\subsection{The datasets}
\label{ssec:dataset}
% ------------------------------------------------
We exploit two types of social network datasets for this work.
The first one is collected by the authors of~\cite{CML11} from Gowalla -- a popular LBSN service back in 2011.
The dataset was collected from February 2009 to October 2010
and it contains 6,442,892 check-ins.
Besides location information, the dataset also includes the corresponding social data
which contains around 1.9 million users and 9.5 million edges.
Due to the large data sparsity,
we mainly focus on the check-in data in two cities in US, 
including New York (NY (G)) and San Francisco (SF (G)).
They are among the areas with most check-ins in the dataset.
In addition, when performing mobility analysis and location prediction, 
we only focus on users who have conducted 
at least 100 check-ins in each city and we term these users as \emph{active users}.

The second dataset is collected from Twitter from December 2014 to April 2015 by the authors of this paper.
Again, we focus on the data in New York (NY (T)) and San Francisco (SF (T)) 
and treat all the geo-tagged tweets (tweets labeled with geographical coordinates) as users' check-ins.
% Our dataset contains two components as well,
% the first one is users' check-in data and the second one is the social graph.
We exploit Twitter's Streaming API\footnote{\url{https://dev.twitter.com/streaming/overview}}
to collect all the geo-tagged tweets.
Each check-in is organized as a 4-tuple.
\[
 \langle {\it uid}, {\it time}, {\it latitude}, {\it longitude} \rangle
\]
Figure~\ref{fig:cisample} depicts a sample of check-ins in New York.
To collect the social relationships among users, we adopt Twitter's REST API\footnote{\url{https://dev.twitter.com/rest/public}}
to query each user's followers and followees.
Two users are considered friends if they follow each other mutually.

Similar to the Gowalla dataset, we only focus on active users (users with more than 100 check-ins) in the Twitter dataset.
Moreover, we also filter out the users who have more than 2,000 check-ins 
since most of them are public accounts such as @NewYorkCP which publishes 16,681 check-ins 
at the exact same location.
Table~\ref{tab:datasets} summarizes the two datasets.
The Twitter dataset is available upon request.

% ==================================================================
\section{Communities}
\label{sec:comdet}
% ==================================================================

We first show how to detect communities in social networks in Section~\ref{ssec:dect}
and then propose a new notion to characterize users' social diversity in Section~\ref{ssec:entropy}.

% ------------------------------------------------
\subsection{Community detection in social networks}
\label{ssec:dect}
% ------------------------------------------------
Community detection in networks (or graphs) has been extensively studied for the past decade
(e.g., see~\cite{N06,RB08,BGLL08,LF10,RB11,ML12,YML13,YL13,MH13,YML14,ML14}).
It has important applications in many fields, including physics, biology, sociology as well as computer science.
The principle behind community detection is to 
partition nodes of a large graph into groups following certain metrics on the graph structure~\cite{LF10}. 
In the context of social networks, besides the social graph, each user is also affiliated with attributes.
These information can also be used to detect communities (e.g., see~\cite{ML12,YML13,ML14}).
For example, people who graduate from the same university can be considered as a community. 
Since the datasets we use only contain social graphs
and no personal information are provided, 
we apply the algorithms that are based on information encoded in graph structure to detect~communities.

According to the comparative analysis~\cite{LF10}, among all the community detection algorithms,
Infomap~\cite{RB08} has the best performance on undirected and unweighted graphs 
and has been widely used in many systems~\cite{NKA14,QSMM15}.
Therefore, we apply it in this work.
Next we give a brief overview of Infomap
and describe how we use it to detect communities.

The main idea of Infomap can be summarized as follows:
information flow in a network can characterize the behavior of the whole network,
which consequently reflects the structure of the network.
A group of nodes among which information flows relatively fast can be considered as one community.
Therefore, Infomap intends to use information flow to detect communities in a network.
In the beginning, Infomap simulates information flow in a network with random walks.
Then the algorithm partitions the network into communities
and exploits Huffman coding to encode the network at two levels.
At the community level, 
the algorithm assigns a unique code for each community 
based on the information flow among different communities;
at the node level, the algorithm assigns a code for each node based on the information flow within the community.
Infomap allows the Huffman codes in different communities (node level) being duplicated 
which results in a more efficient encoding (less description length).
In the end, finding a Huffman code 
to concisely describe the information flow 
%(by regulating the lower bound of code length following Shannon's source coding theorem) 
while minimizing the description length
is thus equivalent to discovering the network's community structure.
In other words, the objective of Infomap is to find a partition of a network 
such that the code length for representing information flow 
among communities and within each community is minimized.
Since it is infeasible to search all possible community partitions, 
Infomap further exploits a deterministic greedy search algorithm~\cite{AMC04,WT07} to find partitions.

\begin{figure}[!h]
\centering
\begin{minipage}[t]{0.42\columnwidth}
   \centering
   \includegraphics[width=\columnwidth]{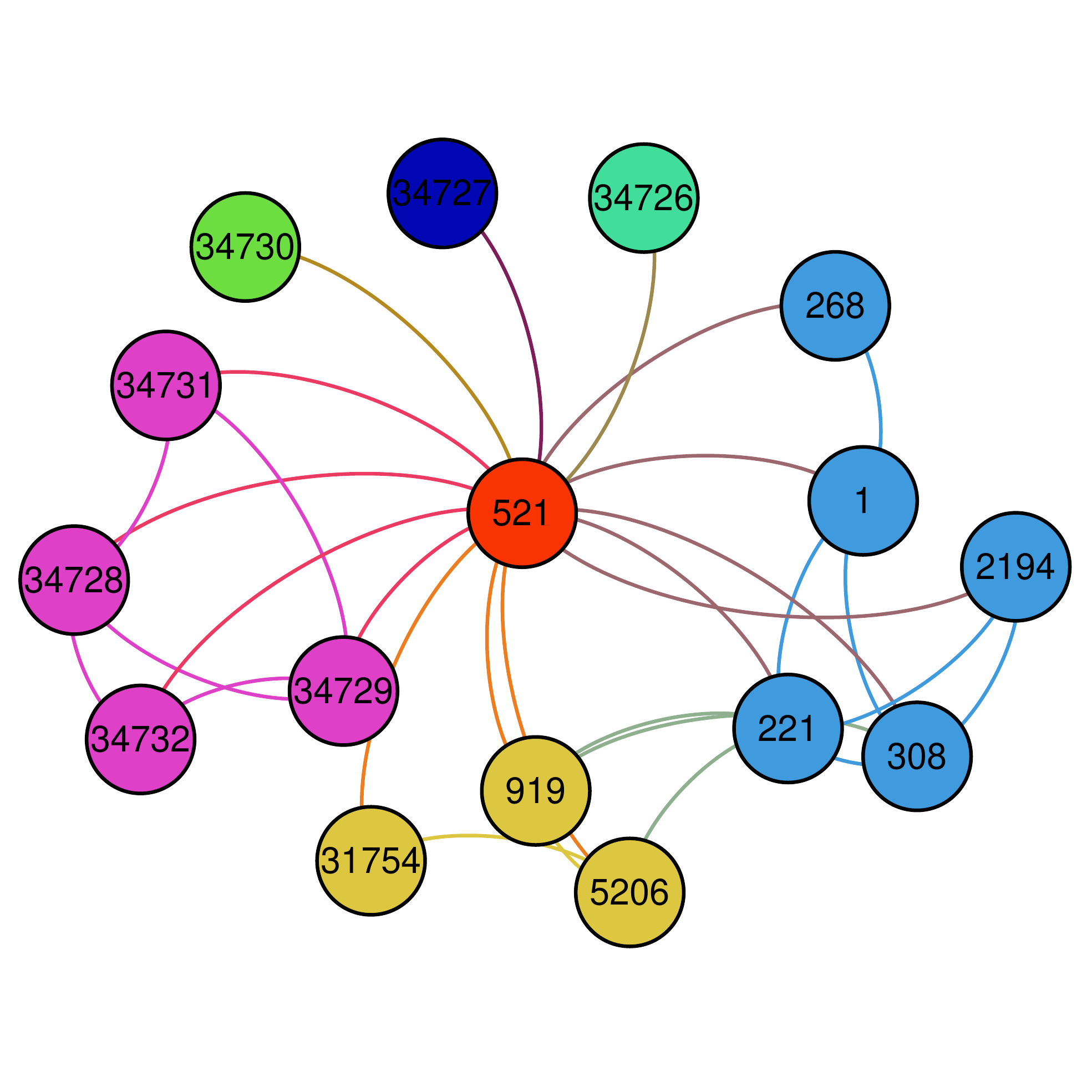}
   \subcaption{User 521 (15 friends)}{\label{fig:comsmall}}
 \end{minipage}
 \vspace{0.001\linewidth}
\begin{minipage}[t]{.56\columnwidth}
  \centering
  \includegraphics[width=\columnwidth]{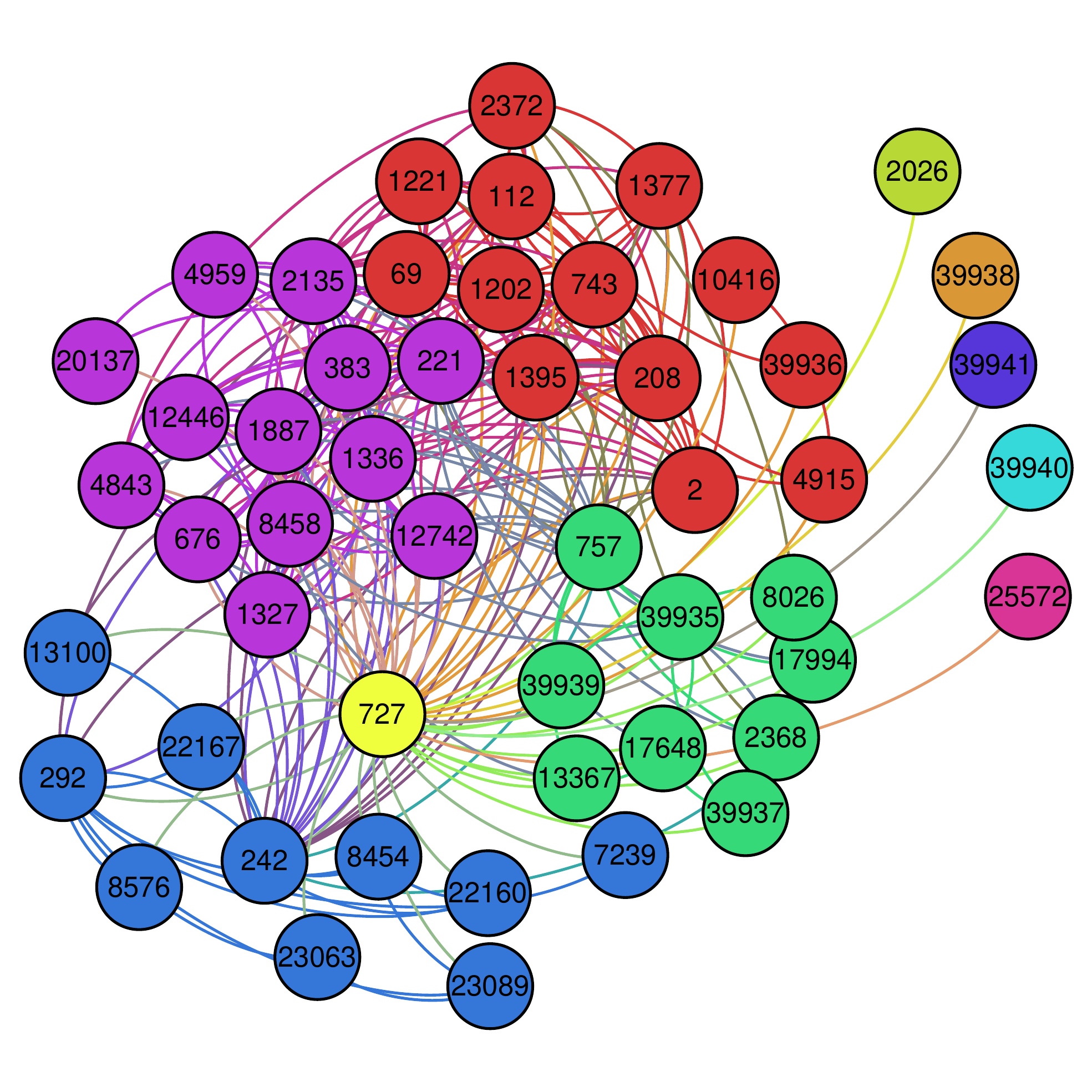}
  \subcaption{User 727 (50 friends)\label{fig:combig}}
\end{minipage}%
\caption{Communities of users 521 and 727.\label{fig:com}}
 \end{figure}

In our work, to detect communities of $\usr$,
we first find all his friends as well as the links among them.
Then, we delete $\usr$ and all edges linked to him
and apply Infomap algorithm to the remaining part of the graph. 
Figure~\ref{fig:com} presents the detected communities of two users in the Gowalla dataset.
Each community is marked with a different color.

%-------------------------------------------------------------
 %-------------------------------------------------------------

\begin{figure}[!t]
\centering
\begin{minipage}[t]{0.49\columnwidth}
   \centering
   \includegraphics[width=1.1\columnwidth]{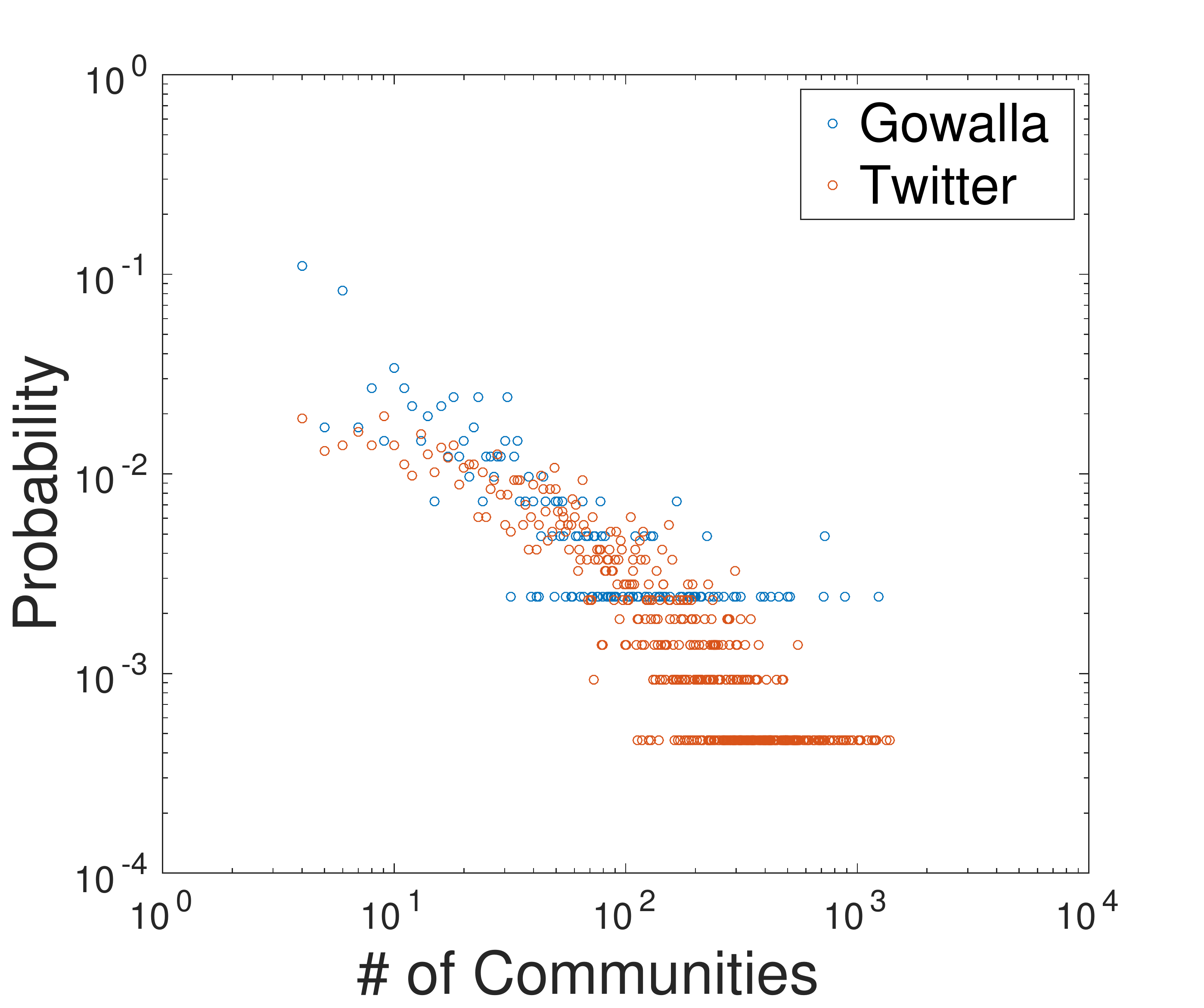}
%\subcaption{Community number \label{fig:distcomnum}}
 \end{minipage}
%       \hspace{0.01\linewidth}
\begin{minipage}[t]{.49\columnwidth}
  \centering
  \includegraphics[width=1.1\columnwidth]{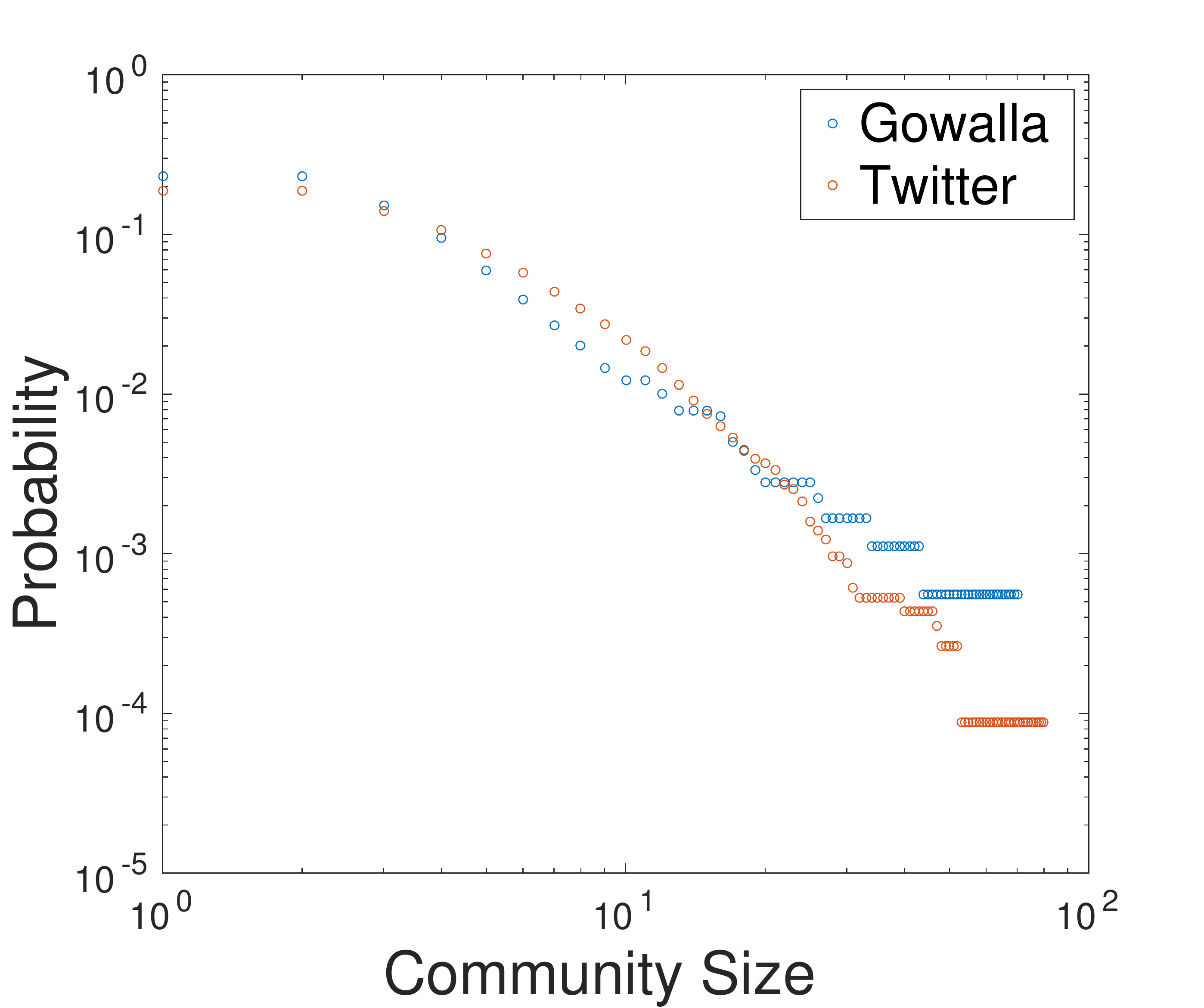}
  %\subcaption{Community size \label{fig:distcomsize}}
\end{minipage}%
\caption{Distribution of users w.r.t\ the number of communities
and distribution of communities w.r.t\ their size.\label{fig:comdist}}
 \end{figure}

\begin{table}[!h]
\centering
\scalebox{1.0}{
\begin{tabular}{| c || r | r| }
\hline
   & Gowalla & Twitter\\
  \hline
  \hline
  Avg.\# of communities & 4.5 & 5.3\\
  \hline
  Avg. community size & 13.2 & 20.8\\  
  \hline
  \end{tabular}
}
\caption{Community summary of active users.\label{tab:com}}
\end{table}
Table~\ref{tab:com} lists the summary of community information of all active users in the two datasets.
Each active user in Gowalla has on average 4.5 communities while the value is 5.3 for the Twitter users.
In addition, the average community size of Twitter users is bigger than Gowalla users (20.8 vs.\ 13.2).
This is because active users in the Twitter dataset 
have more friends than those in the Gowalla dataset (see Table~\ref{tab:datasets}),
which indicates general social network services, such as Twitter, contain more users' social relationships 
than LBSN services, such as Gowalla.
In spite of the differences on the average value in Table~\ref{tab:com},
community number and community size in the two datasets follow a similar distribution.
As we can see from Figure~\ref{fig:comdist},
both community number and size follow the power law:
most of the users have small number of communities and most of the detected communities are small as well.

%-------------------------------------------------------------
%-------------------------------------------------------------

% ------------------------------------------------
\subsection{Community entropy}
\label{ssec:entropy}
% ------------------------------------------------
After detecting communities,
we are given a new domain of attributes on users.
We are particularly interested in
how diverse a user's communities are.
We motivate this \emph{social diversity} through an example.
Suppose that a user is engaged in many communities, 
such as colleagues at work, family members, college friends, chess club, basketball team, etc,
then he is considered an active society member.
Users of this kind are always involving in different social scenarios or environments,
and his daily behaviors are largely dependent on his social relations.

Although we do not have the semantics of each of our detected communities, 
such as the aforementioned colleagues at work or chess club,
we can still use the information encoded in the graph to define a user's social diversity.
For instance, for a user with several communities whose sizes are more or less the same,
his social diversity is for sure higher than those with only one community.

To quantify the social diversity of a user,
we introduce the notion of \emph{community entropy}.

\smallskip
\noindent\textbf{Definition 1.}
For a user $\usr$, his {\it community entropy} is defined as
 \[
  \coment{\usr} =  \frac{1}{1 - \alpha} 
  \ln\sum_{\com\in \Com{\usr}}(\frac{\vert\com \vert}{\vert\fri{\usr}\vert})^{\alpha}.
 \]
Our community entropy follows the definition of R\'{e}nyi entropy~\cite{Ren61}.
Here, $\alpha$ is called the order of diversity,
it can control the impact of community size on the value 
which gives more flexibility to distinguish users when focusing on the sizes of their communities.
In simple terms, our community entropy,
\begin{itemize}  
\item when $\alpha\! >\! 1$, values more on larger communities;
\item when $\alpha\! <\! 1$, values more on smaller communities.
\end{itemize}
The limit of $\coment{\usr}$ with $\alpha \rightarrow 1$ 
is the Shannon entropy.\footnote{\url{https://en.wikipedia.org/wiki/Renyi_entropy}}
In general, if a user has many communities with sizes equally distributed, 
then his community entropy is high and this indicates that his social relations are highly diverse.

\begin{figure}[h]
\centering
\begin{minipage}[t]{0.49\columnwidth}
   \centering
   \includegraphics[width=1.1\columnwidth]{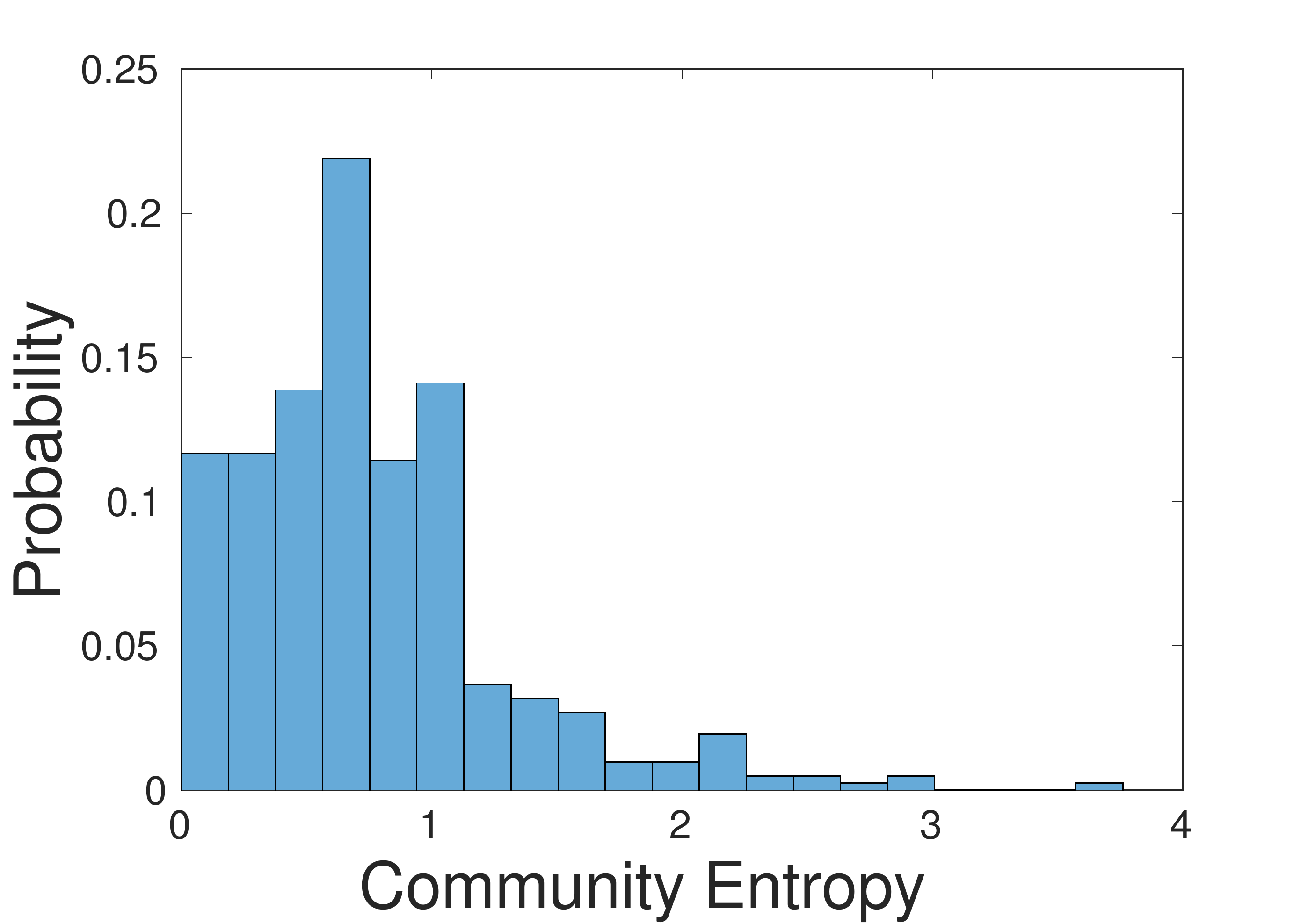}
 \end{minipage}
\begin{minipage}[t]{.49\columnwidth}
  \centering
   \includegraphics[width=1.1\columnwidth]{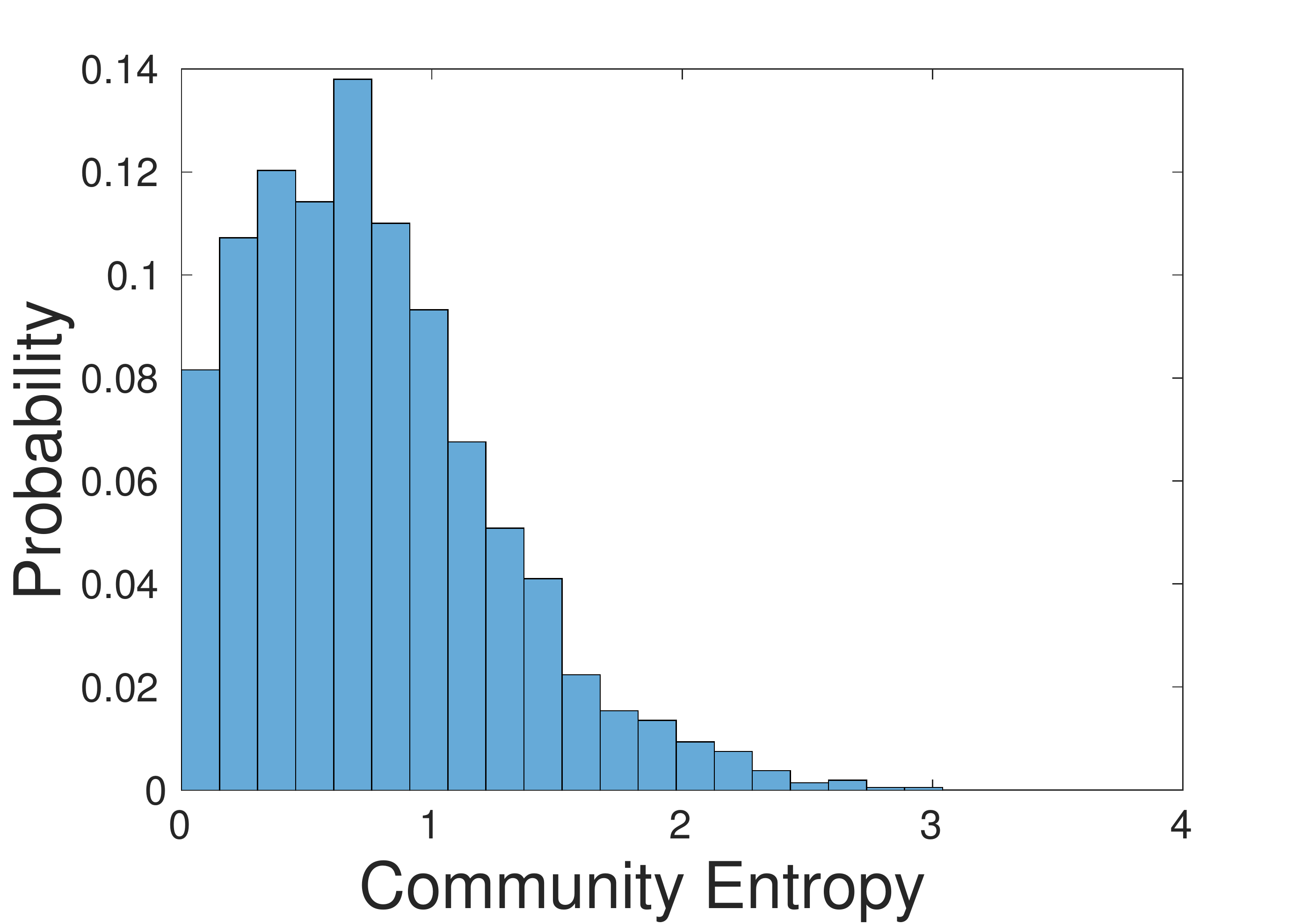}
\end{minipage}%
\caption{Distribution of community entropies of active users in Gowalla (left)
and Twitter (right). \label{fig:distcoment}}
 \end{figure}

We set $\alpha > 1$  in the following discussion
to limit the impact of small communities 
since a user may randomly add strangers as his friends in online social networks 
and these strangers normally form small communities (such as a one-user community\footnote{In our community detection algorithm,
if $\usr'$ himself forms a community of $\usr$, then it indicates that $\usr'$ does not know any other friends of $\usr$.}), 
which have less impact on the user's mobility.
For example, if a user $\usr$ has three communities with sizes equal to 1, 1 and 10,
then his communities are not that diverse following the above intuition.
When we set $\alpha$ less than 1, such as 0.5,
we have $\coment{\usr} = 0.79$ which is a high value indicating $\usr$'s social circles are diverse.
On the other hand, if we set $\alpha$ bigger than 1, such as 10,  $\coment{\usr}$ drops to 0.20 
which captures our intuition.
% Following our intuition, the latter case is more suitable.
In the following experiments, we set $\alpha = 10$ when calculating users' community entropies.
Note that we have also set $\alpha$ to other numbers bigger than one and observed similar results.
Figure~\ref{fig:distcoment} shows the histogram of community entropies of all active users in two datasets.

%-------------------------------------------------------------
%-------------------------------------------------------------
\begin{figure*}[t]
 \centering
 \begin{minipage}[t]{.55\columnwidth}
   \centering
  \includegraphics[width=1\columnwidth]{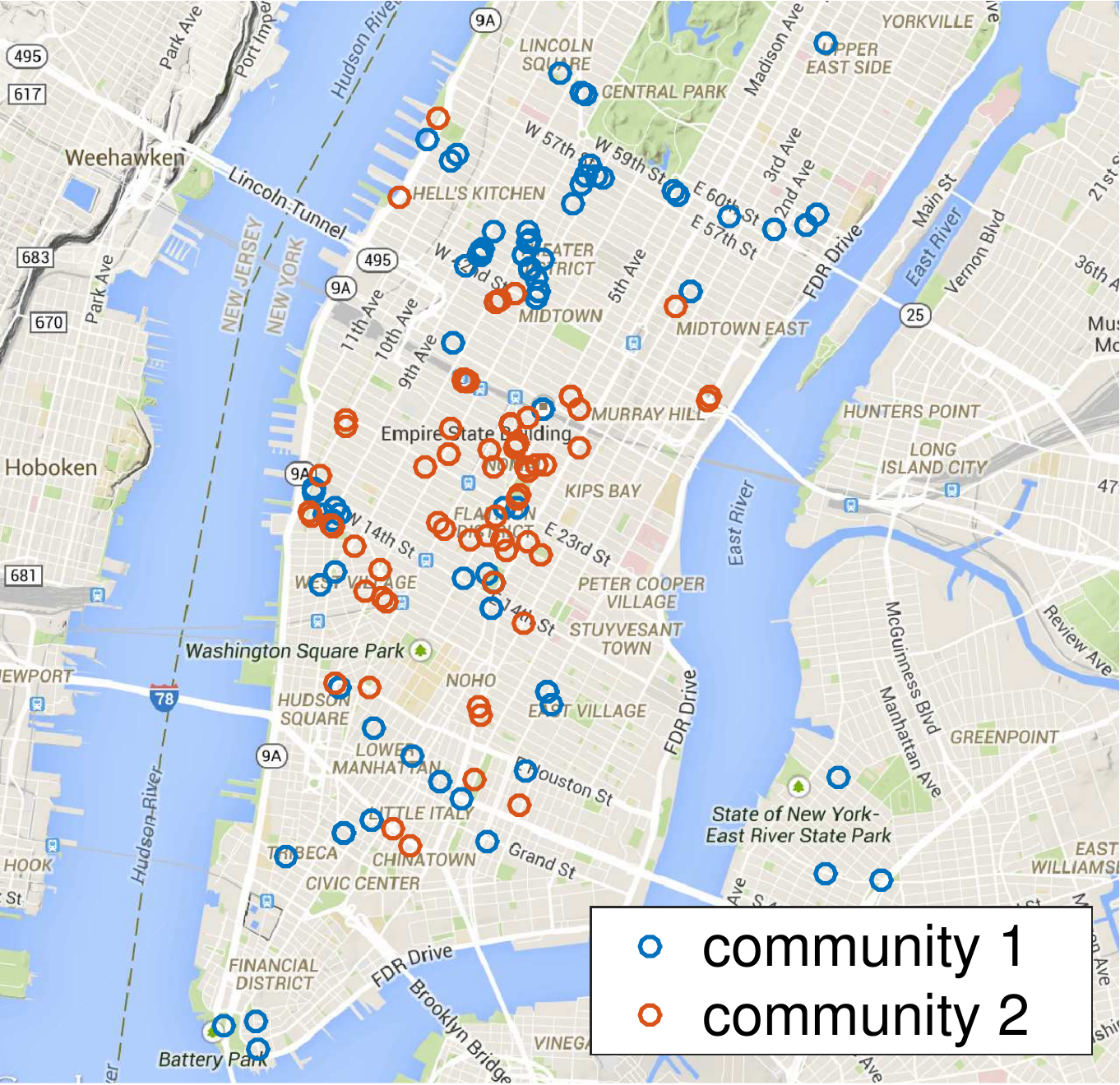}
  \subcaption{\label{fig:caseplot_ny}}
 \end{minipage}
 \begin{minipage}[t]{.74\columnwidth}
   \centering
  \includegraphics[width=1\columnwidth]{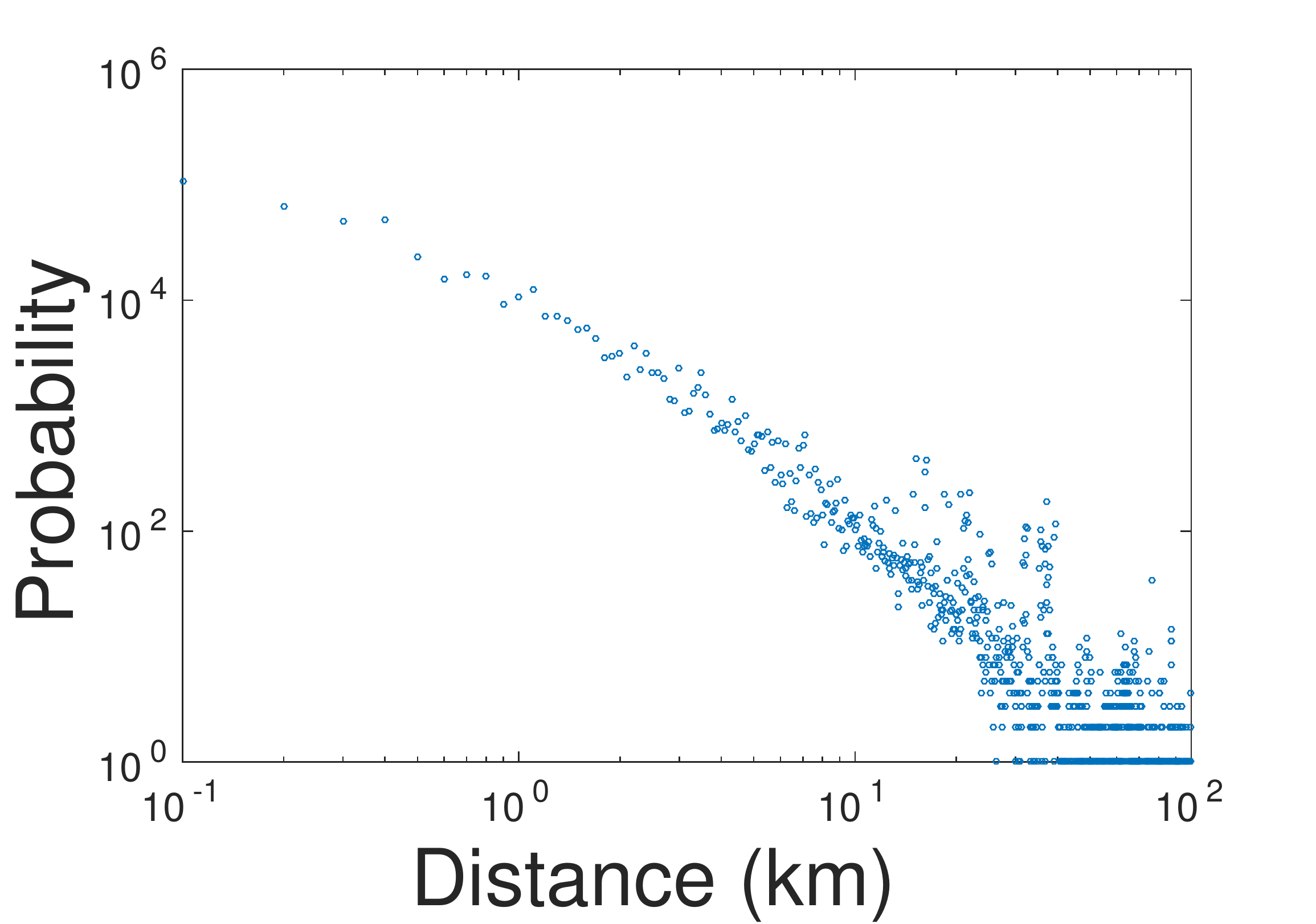}
  \subcaption{\label{fig:distdist}}
 \end{minipage}
 \begin{minipage}[t]{.74\columnwidth}
   \centering
  \includegraphics[width=1\columnwidth]{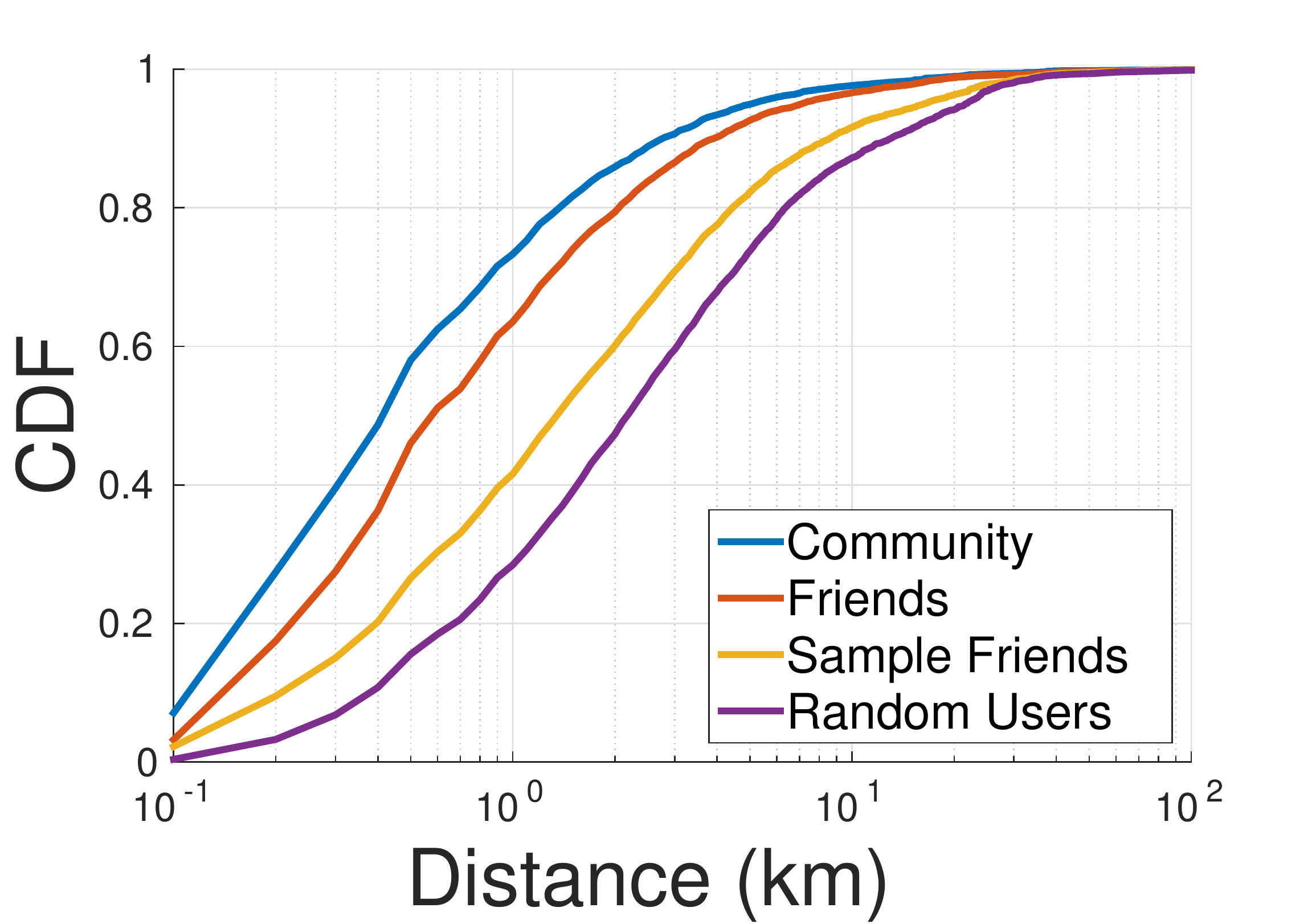}
  \subcaption{\label{fig:compdist}}
 \end{minipage}
 \caption{(a) A user's two communities' check-ins in Manhattan;
 (b) distribution of distances between users and their communities;
 (c) cumulative distribution function of distances between users and their communities, friends, sample friends and random users.}
\end{figure*} 

% ==================================================================
\section{Communities and Mobility}
\label{sec:commob}
% ==================================================================

It has been proved that social factors play an important role on users' mobility, e.g., see~\cite{CML11}. 
For instance, one may go to lunch with his friends or go to a bar to hangout with his friends.
Meanwhile, for a user, friends of his social networks (as well as in real life) are not all equal. 
Instead friends normally belong to certain communities.
When considering a user's mobility, 
intuitively different communities can impose different influence within certain contexts or social environments. 
Continuing with the above example, 
the people the user has lunch with are normally his colleagues 
while the people he meets at night are his close friends.
Therefore, in order to analyze the impact from a user's social relations on his mobility,
it is reasonable to focus on social influence at the community level.

In this section, we first study communities' influence on users' mobility.
After that, we study the characteristics of the influential communities 
with the following two intuitions in mind:
(1) a user's daily activities are constrained, 
and the number of communities he interacts with is limited;
(2) communities influence a user's social behavior under different contexts.

 %-------------------------------------------------------------
\begin{figure*}[!t]
 \centering
 \begin{minipage}[t]{.68\columnwidth}
   \centering
   \includegraphics[width=1.1\columnwidth]{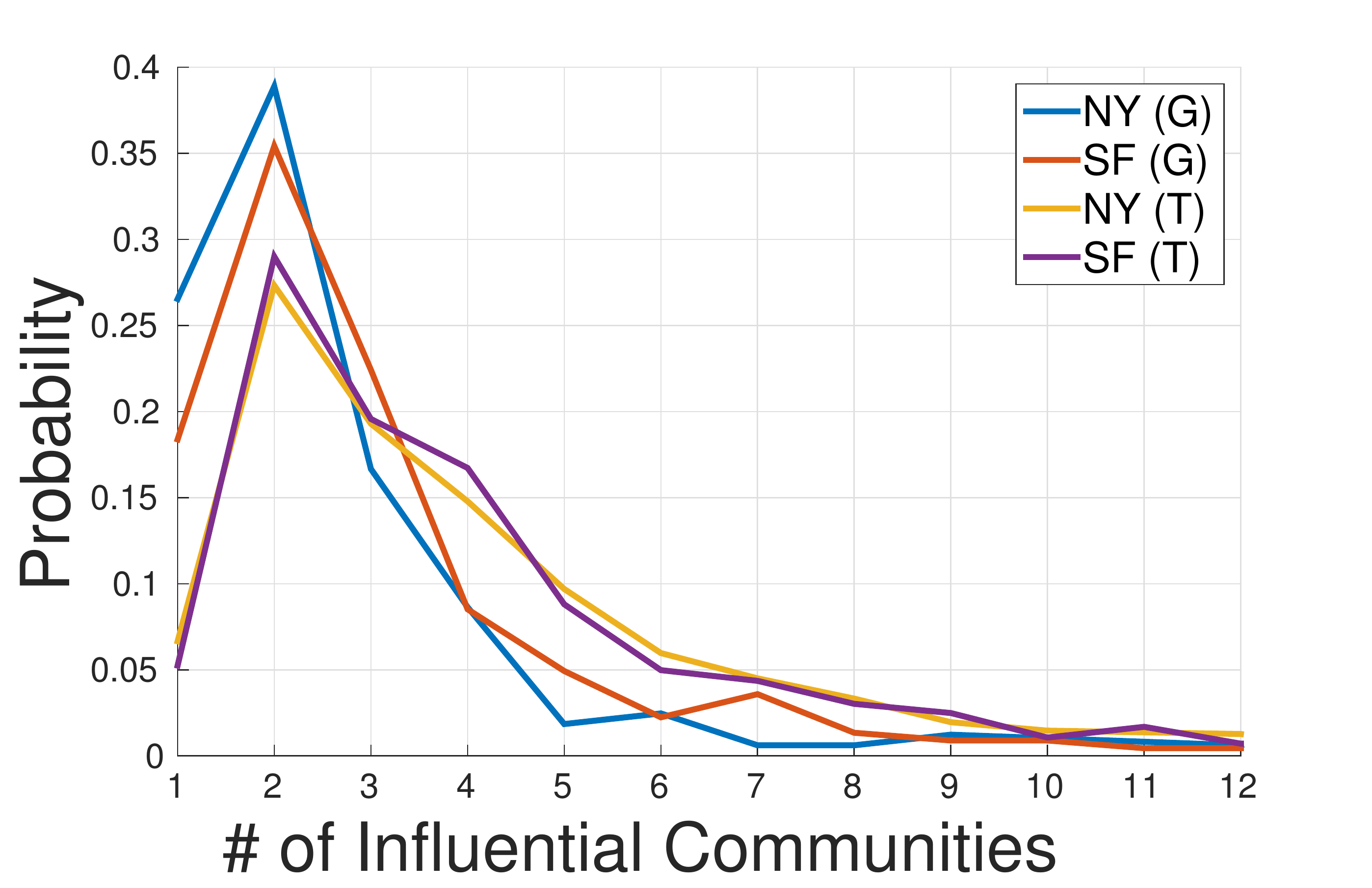}
    \subcaption{\label{fig:distinfcom}}
 \end{minipage}
 \begin{minipage}[t]{.68\columnwidth}
   \centering
  \includegraphics[width=1.1\columnwidth]{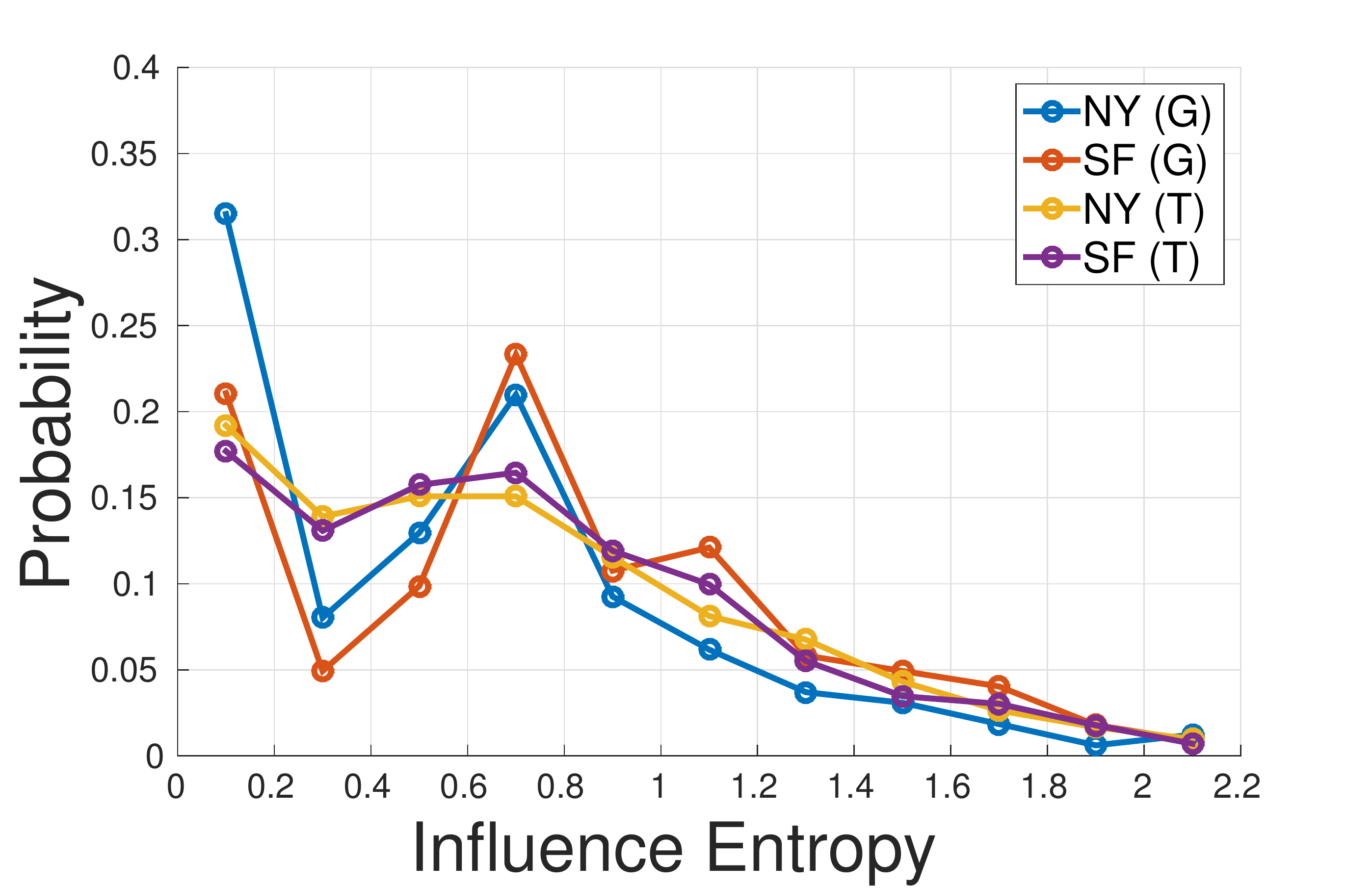}
  \subcaption{\label{fig:distinfent}}
 \end{minipage}
 \begin{minipage}[t]{.68\columnwidth}
   \centering
  \includegraphics[width=1.1\columnwidth]{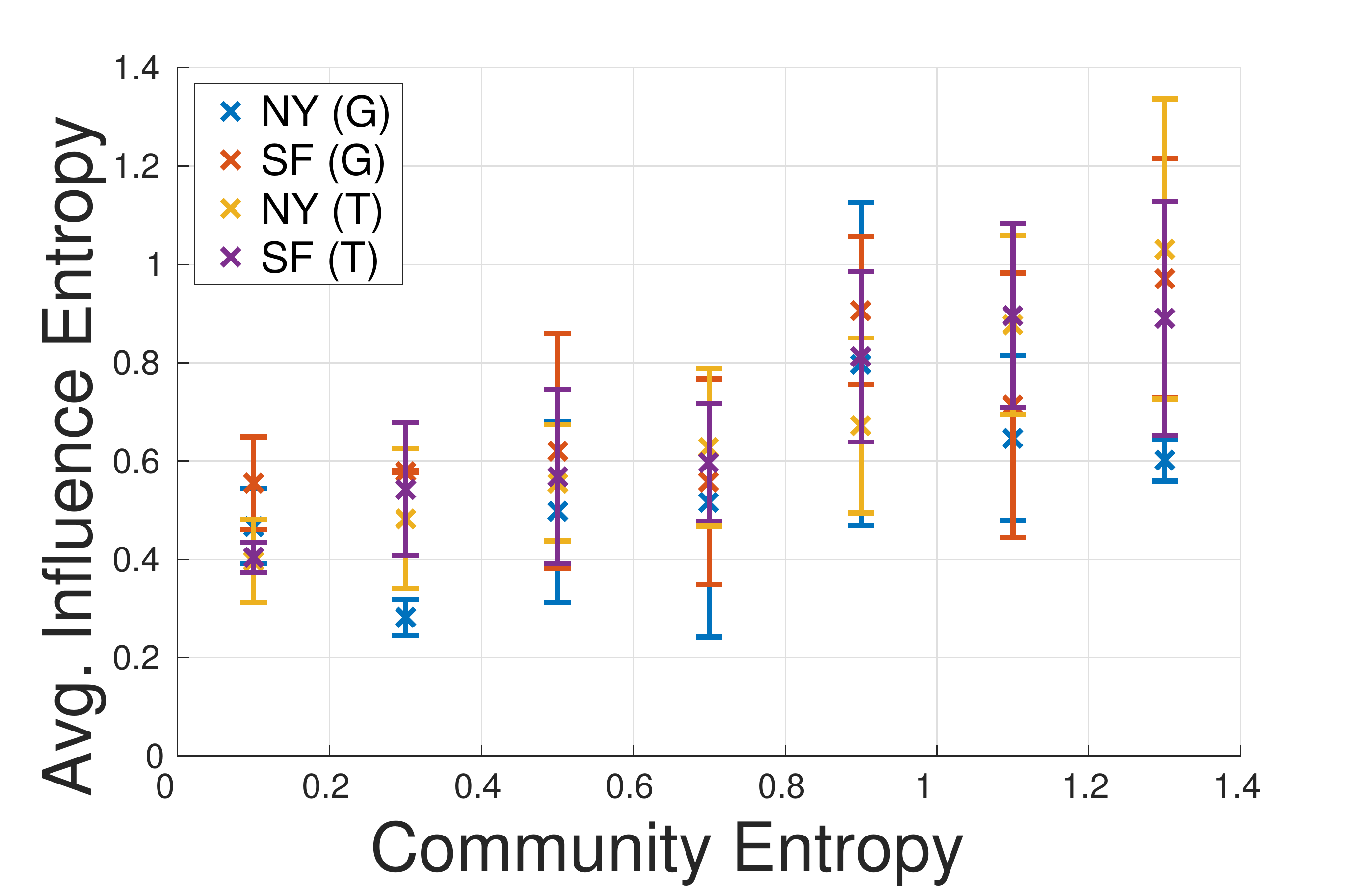}
  \subcaption{\label{fig:infcoment}}
 \end{minipage}

 \caption{(a) Distribution of the number of influential communities;
 (b) distribution of influence entropies (bucketed by 0.2)
 (c) influence entropy vs. community entropy.}
\end{figure*} 
 %-------------------------------------------------------------

% ----------------------------------------------
\subsection{Influential communities}
% ----------------------------------------------

Figure~\ref{fig:caseplot_ny} depicts a user's two communities' check-ins in Manhattan of the New York City.
We can observe a quite clear separation between these two communities' check-ins:
members of community~1 mainly visit Uptown and Midtown Manhattan
while community~2 focuses more on Midtown.
This indicates that different communities have their social activities at different areas.
In a broader view, this shows that partitioning users' check-ins at the social network level (through community detection)
can result in meaningful spatial clusters as well.

A single community also has several favorite places.
For example, community 1 in Figure~\ref{fig:caseplot_ny} 
visits Times Square and Broadway quite often
while members of community 2 like to stay close to Madison square park.
A user may socialize with different communities at different places,
for example, he may go to watch a basketball game with his family at the stadium 
and have lunch with his colleagues near his office.
Therefore, to study influences on mobility from communities to a user,
we need to summarize each community's \emph{frequent movement areas}.
To discover a community's frequent movement areas, 
we perform clustering on all locations that the community members have been to.
Each cluster is then represented by its central point 
and a community's frequent movement areas are thus represented by the centroids of all clusters.
% Formally, a community $\com$'s
% frequent movement areas is defined as
% \[
%   fm(\com) = \{fm_1, \ldots, fm_n\}
% \]
% where $fm_i$ is the centroid of the community's $i$th cluster.
The clustering algorithm we use is the agglomerative hierarchical clustering.
We regulate that any two clusters can be aligned only if the distance between
their corresponding centroids is less than 500m
which is a reasonable range for human mobility.

To illustrate the mobility influence from communities to users,
we choose to use `distances'.
More precisely, we represent the influence
by the distances between a user's locations 
and the frequent movement areas of his communities.
Shorter distances imply stronger influences.
For each location a user has visited,
we calculate the distances between the location and all his communities' frequent movement areas.
Then, for each community of the user,
we choose the shortest distance 
between the location and the community's frequent movement areas
as the \emph{distance} between the location and the community.
The community which has the smallest distance
to the location is considered as
the \emph{influential community} of the user at this location.
The distance between the influential community and the user's location
is further defined as the distance between the user's location and his communities.
% Formally, for a location $\ell$ of $\usr$, 
% we use ${\it infcom}(\usr, \ell)$ to denote $\usr$'s influential community at $\ell$
% and ${\it infdist}(\usr,\ell)$ to denote the minimal distance between $\ell$
% and ${\it infcom}(\usr, \ell)$'s frequent movement areas.
% In the end, the distance between a user $\usr$ and his communities,
% i.e., the influence $\usr$ gets from his communities, is defined as
% \[
%  {\it inf}(\usr) = \{{\it infdist(\usr, \ell)}\ \vert\ \ell\in \Ci{\usr}\}
% \]
% where we abuse the notation $\ell\in \Ci{\usr}$ 
% to represent the location of $\usr$'s one check-in.
% \JP{I am not comfort with the above notations, which only appear here and are never used later.}
% \YZ{I am also not comfort with it, i add it here only because the third viewer asked us to add some math here,
% this part in fact is not well linked with other parts of the paper, so I comment it out directly}
Note that a user can have multiple influential communities
and an influential community can influence a user on multiple locations.

Figure~\ref{fig:distdist} depicts the distribution of distances 
between users' locations and their communities in New York and San Francisco in the two datasets.
As we can see, most of the distances are short
which indicates the communities are quite close to users' locations.
To illustrate that these short distances are not due to the limits of the city areas,
for each location of a user, we pick some random users in the city,
summarize their frequent movement areas through clustering
and find the minimal distance between their frequent movement areas and the location.
In Figure~\ref{fig:compdist}\footnote{The results in Figure~\ref{fig:compdist} 
are based on the data from two cities in both datasets.}, 
the curve of cumulative distribution function (CDF) for these random users (purple) 
is much lower than the one for communities (blue).
This means that these random users are farther away from the users than communities.
To show that community is a meaningful level to study mobility,
we also calculate distances between a user and all his friends.
The curve for friends (red) in Figure~\ref{fig:compdist} is lower than the one for communities as well,
meaning that a user is closer to his communities than to all his friends in general.
As a user's community is a subset of his friends,
to illustrate that the shorter distances for communities than friends
are not caused by frequent movement areas clustered from a small number of friends' check-ins,
for each community of a user,
we randomly sample the same number of his friends to build a ``virtual'' community
and calculate the distances between the user and his virtual communities.
The CDF curve in Figure~\ref{fig:compdist} (yellow) shows that these virtual communities 
are even farther away from users than all friends. 

From the above analysis, 
we conclude that (1) communities have strong influences on users' mobility
and (2) community is a meaningful resolution to study users' mobility.

% ----------------------------------------------
\subsection{Number of influential communities}
% ----------------------------------------------
Research shows that a user's mobility is constrained geographically (see~\cite{CCLS11, CML11}), 
e.g., a user normally travels in or around the city where he lives.
Meanwhile, social relations are not restricted by geographic constrains.
For instance, a user's college friends as a community can spread all over the world.
Now we focus on how many communities actually influence a user's mobility
i.e., how many influential communities a user has.
Intuitively, this number should be small as each user only interacts 
with a limited number of communities in his daily life such as colleagues and family.

We plot the distribution of 
the number of user's influential communities in Figure~\ref{fig:distinfcom}.
From two datasets, we can observe a similar result.
Most of the users are influenced only by a small number of communities
and there are more users who have two influential communities than others.
For example, almost 30\% of users in
New York have two influential communities in the Twitter dataset.

Each location corresponds to an influential community.
We proceed with studying how a user's influential communities 
are distributed over his check-ins.
We first propose a notion named \emph{influence entropy}, it is defined as
\[
{\it infent}(\usr) = -\sum_{c\in \Com{\usr}}\frac{\vert \Ci{\usr, \com}\vert}{\vert\Ci{\usr}\vert}\ln\frac{\vert\Ci{\usr, \com}\vert}{\vert\Ci{\usr}\vert}
\]
where $\Ci{\usr, \com}$ represents $\usr$'s check-ins that are closest to the community $\com$.
The influence entropy is defined in the form of Shannon entropy:
higher influence entropy indicates that the user's locations 
are close to his different communities more uniformly.
Figure~\ref{fig:distinfent} depicts the distribution of users' influence entropies.
As we can see, in New York (NY (T)), around 20\% of users' influence entropies are between 0 and 0.2
which means they have one dominating influential community 
that is close to most of their locations.
We also notice that there is a peak around 0.6 in all the cities.
For example, if a user $\usr$'s 50\% check-ins corresponds to one influential community and 
the other 50\% corresponds to another one,
then ${\it infent}(\usr) = 0.69$ which falls into this range.
This shows that around 20\% of users are influenced by their two major communities at a similar level.

Community entropy introduced in Section~\ref{sec:comdet} is a notion for capturing a user's social diversity.
We further study the relationship between community entropy and influence entropy.
As shown in Figure~\ref{fig:infcoment},
more diverse a user's social relationship is,
more probably his locations are distributed uniformly over his influential communities.

From the above analysis,
we conclude that only a small number of communities have influences on users' mobility.

% ----------------------------------------------
\subsection{Communities under contexts}
% ----------------------------------------------

Influential communities are constrained by contexts.
For instance, a user has lunch with his colleagues 
and spends time with his family near where he lives.
Here, the lunch hour and the home location can be considered as social contexts,
and the two communities (colleague and family) have impact on the user's behavior
under each of the context, respectively.
Thus it is interesting to study whether this hypothesis holds generally.

 %-------------------------------------------------------------
\begin{figure*}[!t]
 \centering
 \begin{minipage}[t]{.5\columnwidth}
   \centering
   \includegraphics[width=1.1\columnwidth]{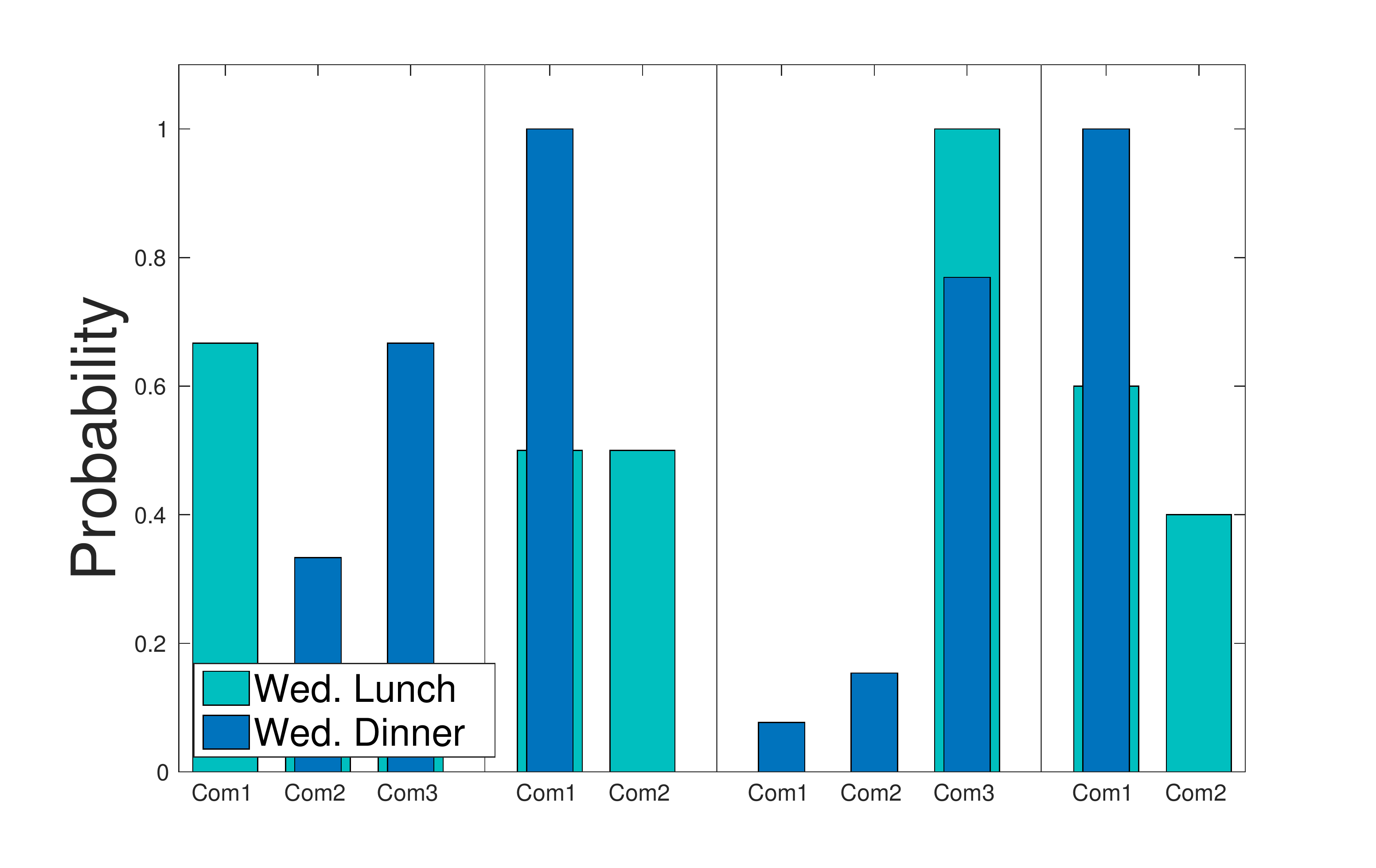}
   \subcaption{NY (G)}
 \end{minipage}
 \begin{minipage}[t]{.5\columnwidth}
   \centering
   \includegraphics[width=1.1\columnwidth]{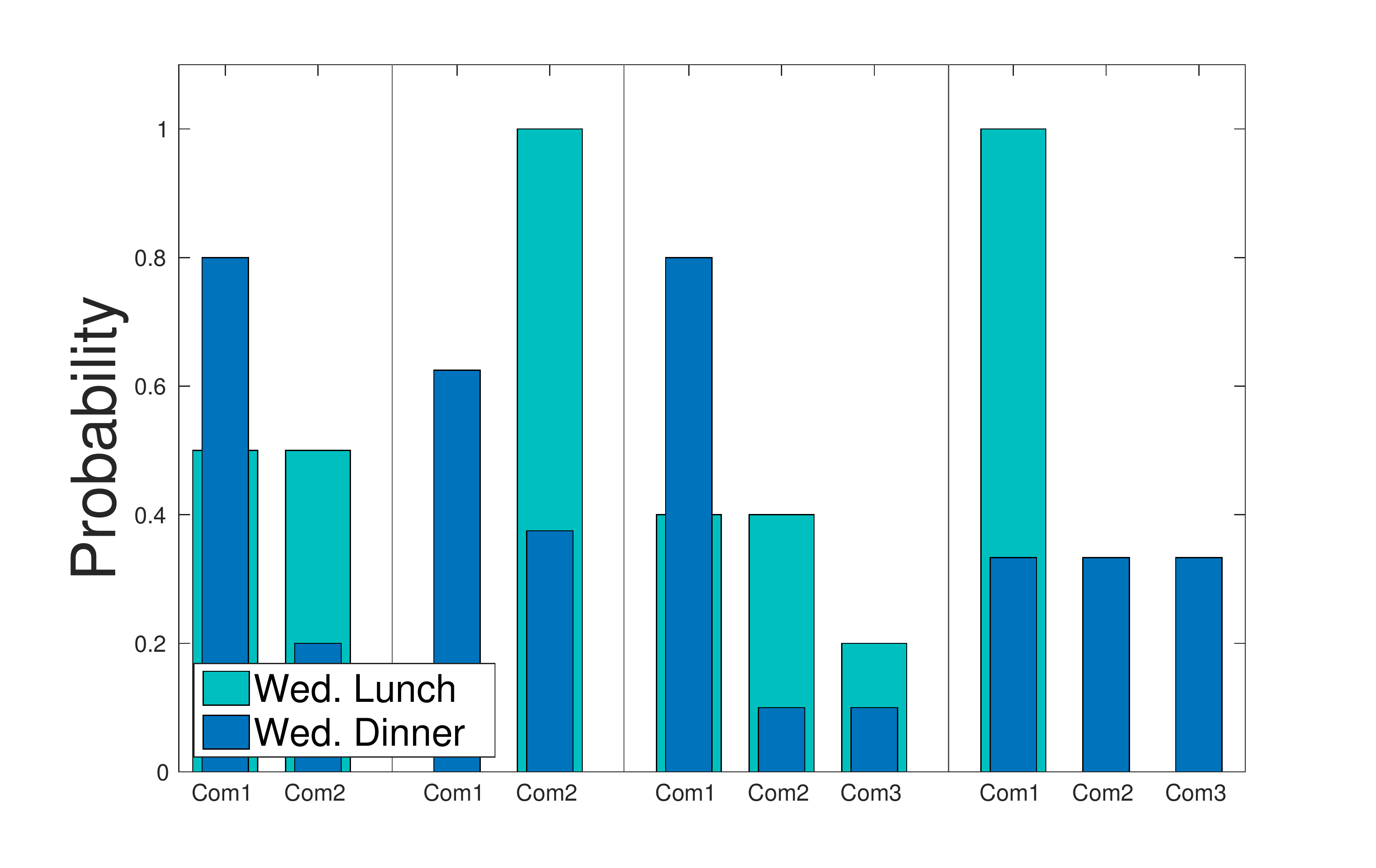}
   \subcaption{SF (G)}
 \end{minipage}
 \begin{minipage}[t]{.5\columnwidth}
   \centering
   \includegraphics[width=1.1\columnwidth]{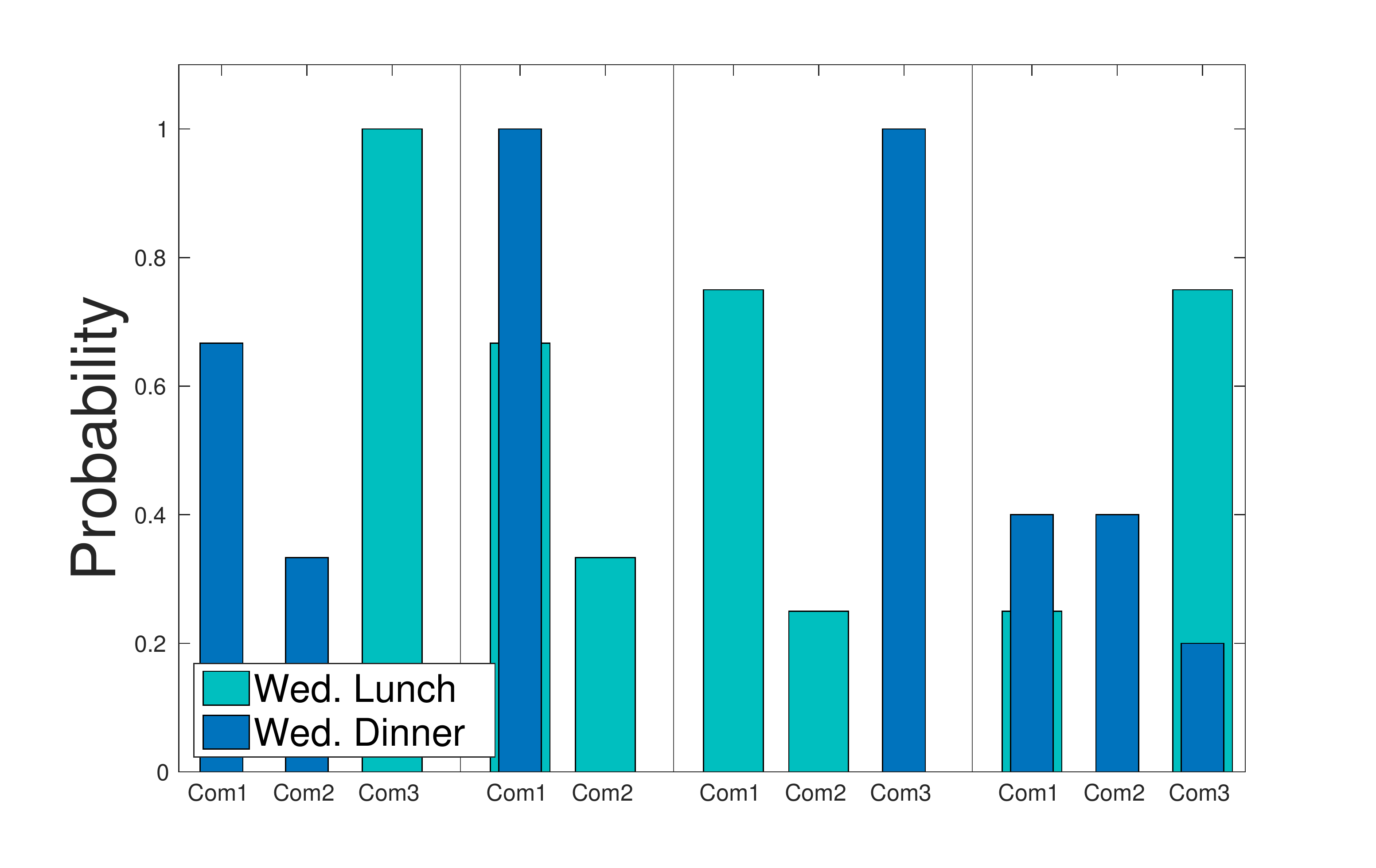}
   \subcaption{NY (T)}
 \end{minipage}
 \begin{minipage}[t]{.5\columnwidth}
   \centering
   \includegraphics[width=1.1\columnwidth]{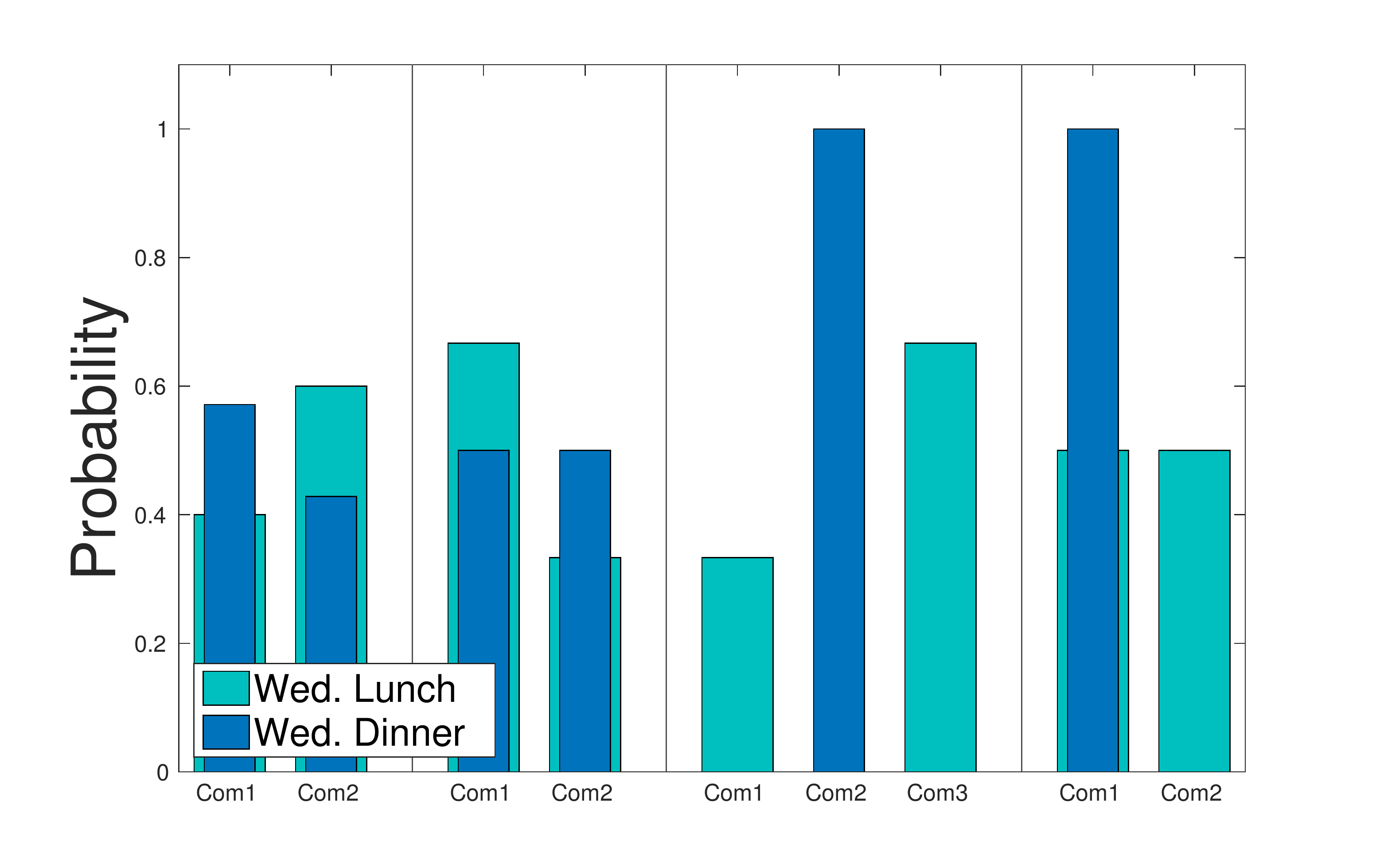}
   \subcaption{SF (T)}
 \end{minipage}
 \caption{Distribution of influential communities on users' check-ins (temporal contexts).\label{fig:timecontext}}
\end{figure*} 
 %-------------------------------------------------------------
 %-------------------------------------------------------------
\begin{figure*}[!ht]
 \centering
 \begin{minipage}[t]{.5\columnwidth}
   \centering
   \includegraphics[width=1.1\columnwidth]{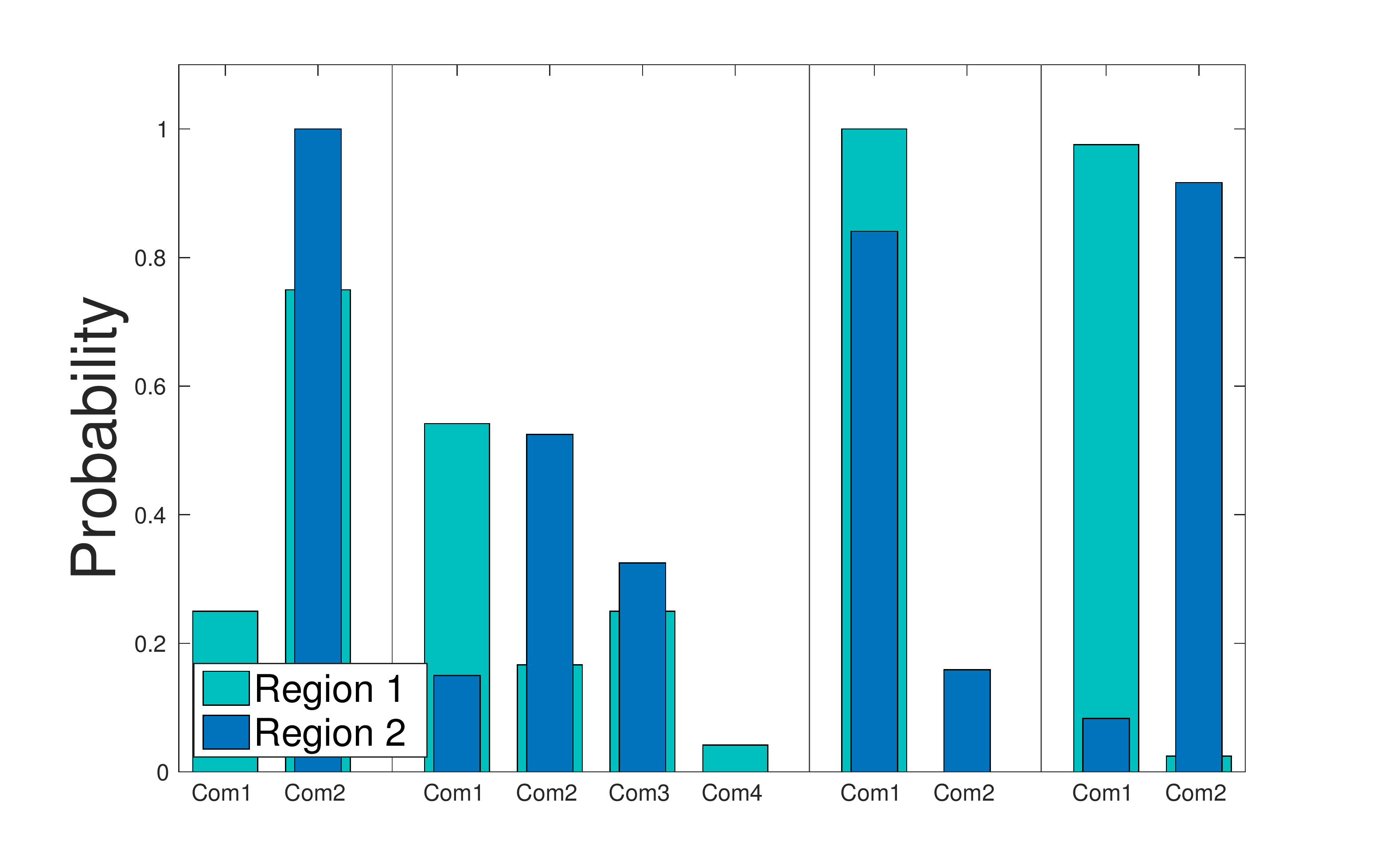}
   \subcaption{NY (G)}
 \end{minipage}
 \begin{minipage}[t]{.5\columnwidth}
   \centering
   \includegraphics[width=1.1\columnwidth]{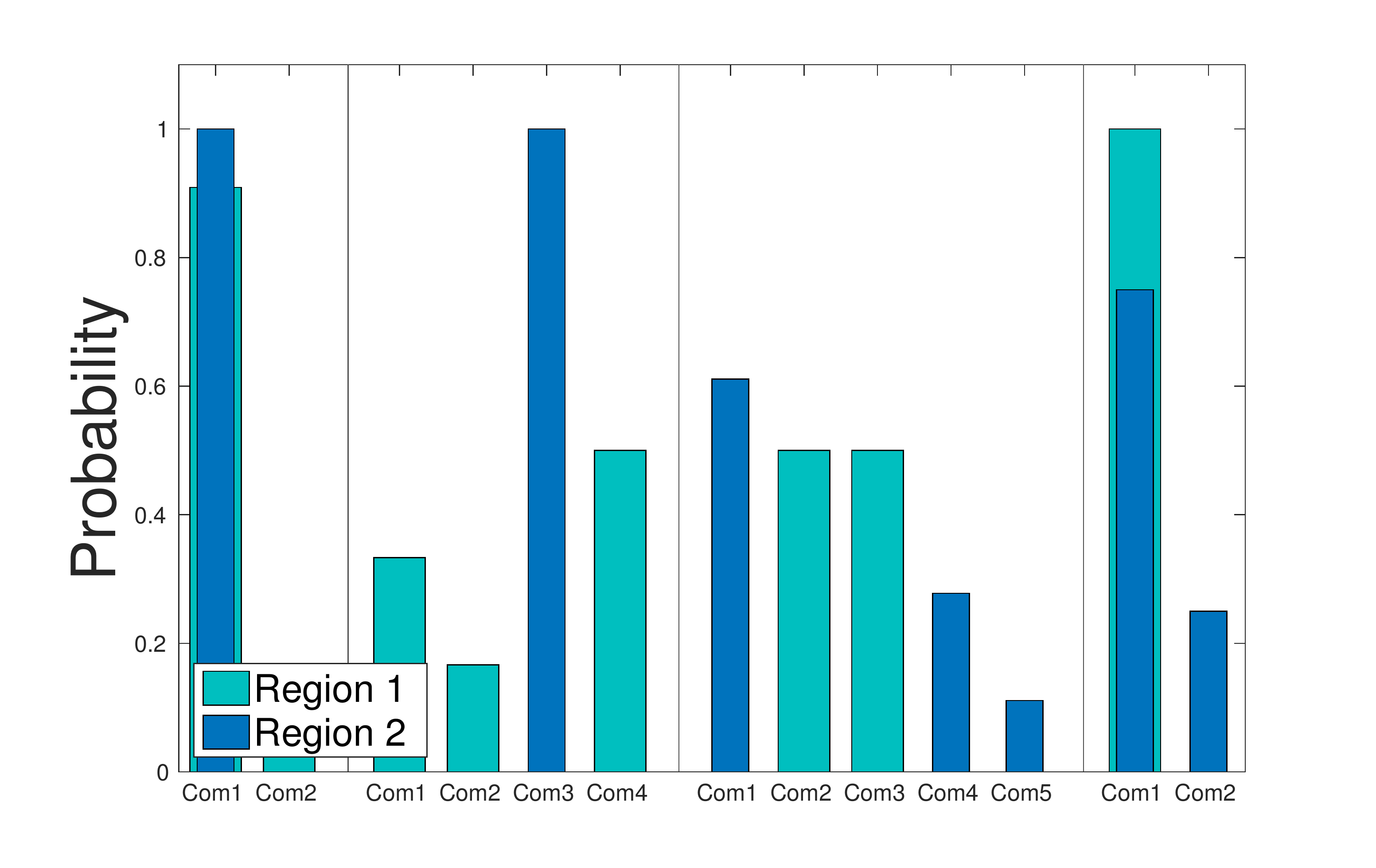}
   \subcaption{SF (G)}
 \end{minipage}
 \begin{minipage}[t]{.5\columnwidth}
   \centering
   \includegraphics[width=1.1\columnwidth]{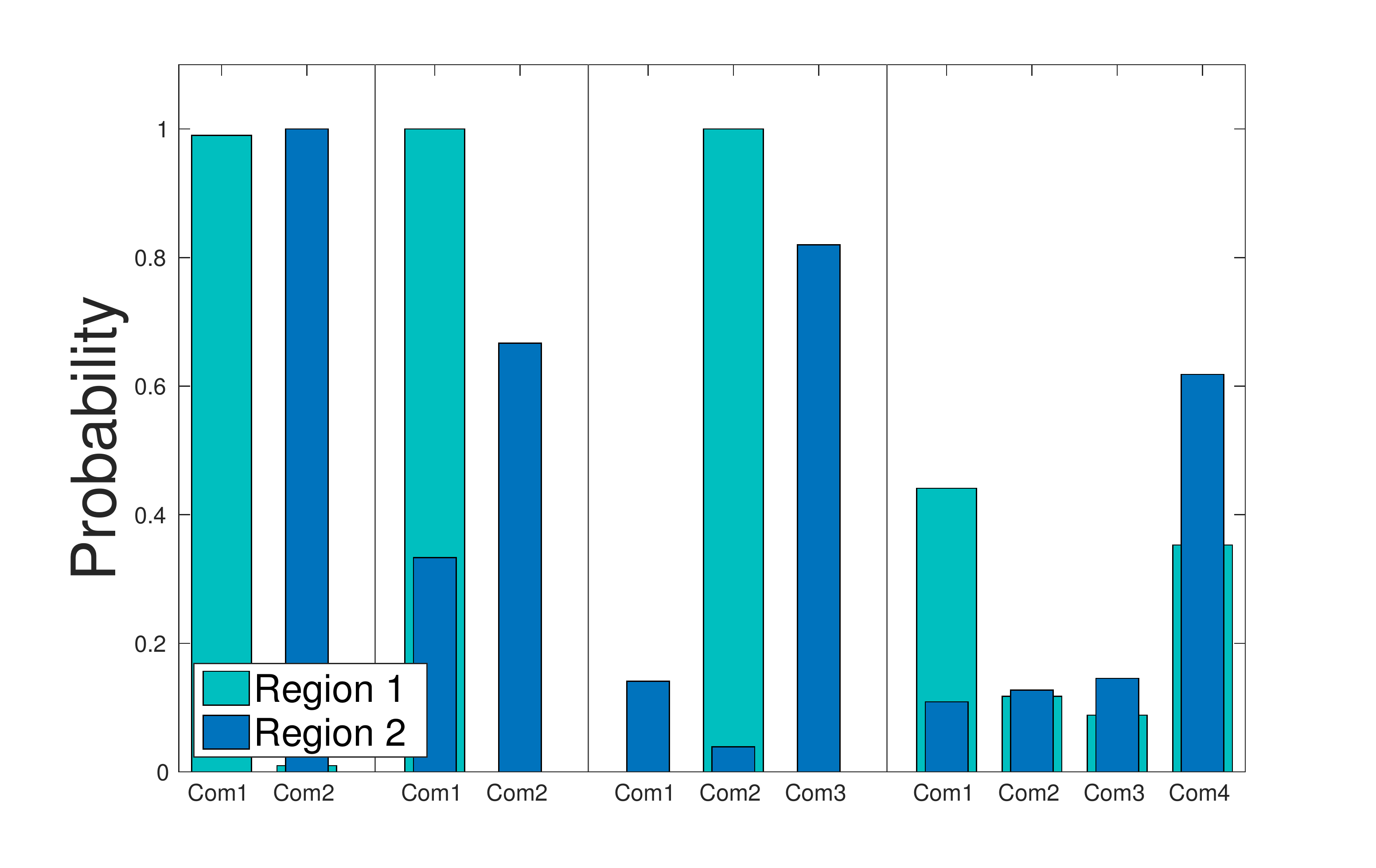}
   \subcaption{NY (T)}
 \end{minipage}
 \begin{minipage}[t]{.5\columnwidth}
   \centering
   \includegraphics[width=1.1\columnwidth]{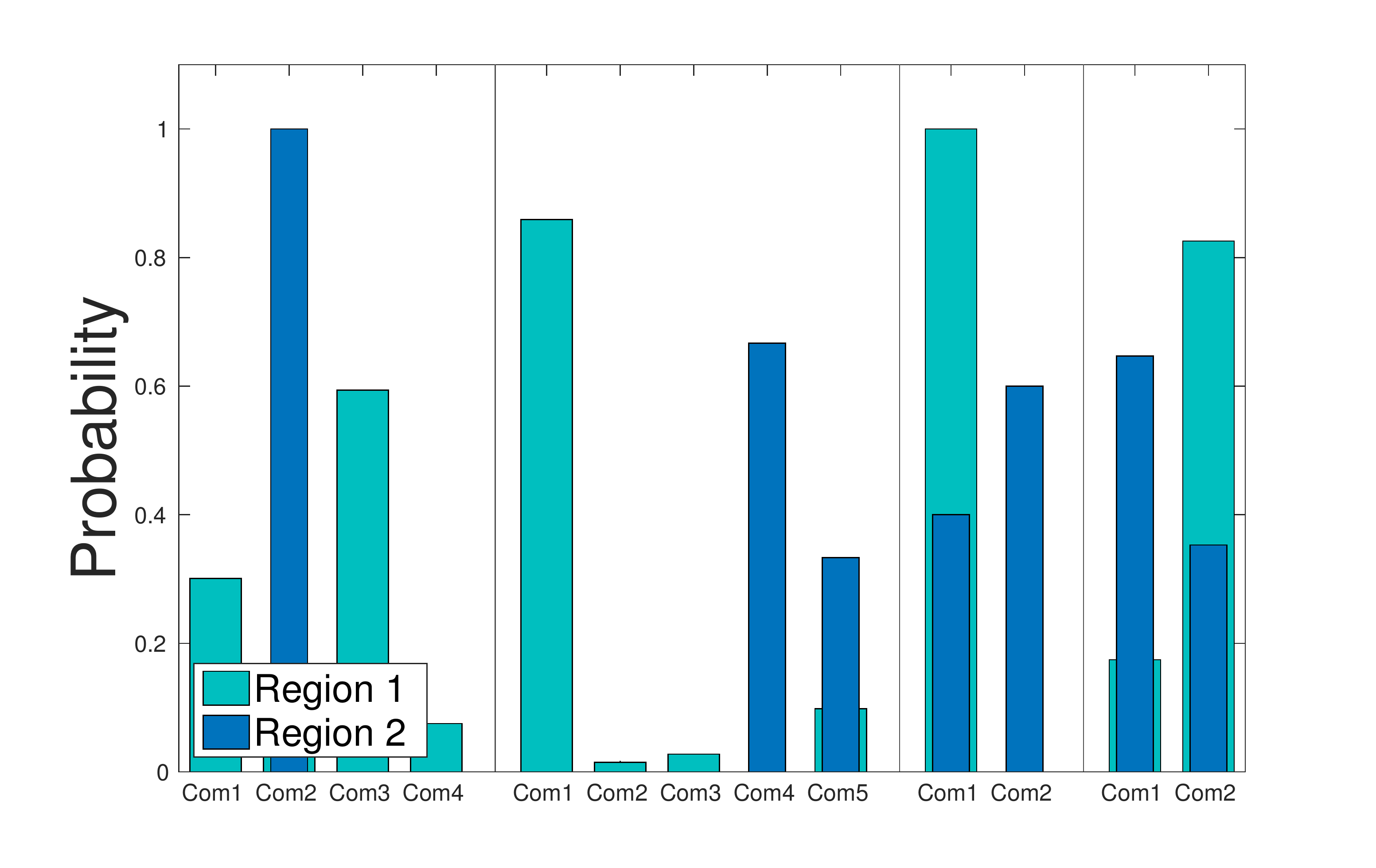}
   \subcaption{SF (T)}
 \end{minipage}
 \caption{Distribution of influential communities on users' check-ins (spatial contexts).\label{fig:loccontext}}
\end{figure*} 
 %-------------------------------------------------------------

\medskip
\noindent\textbf{Temporal contexts.}
First, we focus on temporal contexts.
The pair of contexts we choose are 
\emph{Lunch} (11am--1pm) and \emph{Dinner} (7pm--9pm) hours on Wednesday.
For each user, we extract his check-ins during lunch and dinner time
and find his influential communities w.r.t.\ these two contexts.
We randomly choose four users and plot the distributions of their check-ins 
over their influential communities under these two contexts in Figure~\ref{fig:timecontext}.
As we can see, a user's communities behave quite differently on influencing his check-ins 
during lunch and dinner time.
For example, the first user in New York in the Twitter dataset is only influenced by his community 3 during lunch time 
while communities 1 and 2 give him similar influences during dinner time.
This simply reflects the fact that the people who users have lunch and dinner with are different.
In addition, users' average influence entropies drop as well under different temporal contexts 
compared with the general case (see~Table~\ref{tab:infentropy_context}),
this suggests that the influential communities tend to become more unique.

For each user during lunch (dinner) time, 
we create a vector where the $i$-th component counts the number of locations that are the closest to community $i$.
We then exploit the cosine similarity between a user's lunch and dinner vectors
as his \emph{influence similarity}.
The results are listed in Table~\ref{tab:infcosim}.
Note that, we also choose other pairs of temporal contexts for analysis, 
such as working hours (9am--6pm) and nightlife (10pm--6am)
and have similar observations.

\medskip
\noindent\textbf{Spatial contexts.}
Next we study the influence of spatial contexts.
In each city, we pick two disjoint regions (called \emph{Region 1} and \emph{Region 2}, respectively) 
including Uptown and Downtown Manhattan in New York and
Golden Gate Park and Berkeley in San Francisco.
Then, we extract users' check-ins in these areas.
By performing the same analysis as the one for temporal contexts,
we observe similar results (see Figure~\ref{fig:loccontext}, Table~\ref{tab:infentropy_context} and Table~\ref{tab:infcosim}).
% For a case study, user 2746's mobility is influenced by community 1 in Uptown 
% and by community 2 in Midtown Manhattan (see Figure~\ref{fig:caseplot_nyclu}).
Note that we choose the areas without 
special semantics in mind, e.g., business areas or residential areas.

\begin{table}[!ht]
\centering
\scalebox{1.0}{
\def\arraystretch{1}
\setlength\tabcolsep{0.7mm}
\begin{tabular}{| c || c | c || c | c|}
  \hline
   Influence entropy    & NY (G) & SF (G) & NY (T) & SF (T)\\
  \hline
  \hline
  General & 0.56 & 0.73 & 0.69 & 0.70 \\
  \hline
  Temporal (\emph{Lunch}) & 0.35 & 0.39 & 0.22 & 0.25\\
  \hline
  Temporal (\emph{Dinner}) & 0.27 & 0.43 & 0.30 & 0.31\\
  \hline
  \hline
  Spatial (\emph{Region 1}) & 0.45 & 0.20 & 0.52 & 0.23\\
  \hline
  Spatial (\emph{Region 2}) & 0.42 & 0.21 & 0.61 & 0.26\\
  \hline
  \end{tabular}
}
\caption{Influence entropy under different social contexts.}
\label{tab:infentropy_context}
\end{table}

\begin{table}[!ht]
\centering
\scalebox{1.0}{
\def\arraystretch{1}
\setlength\tabcolsep{0.7mm}
\begin{tabular}{| c || c | c || c | c |}
  \hline
  Influence similarity &NY (G) & SF (G) & NY (T) & SF (T)\\
  \hline
  \hline
   Temporal & 0.80 & 0.74  & 0.67 & 0.66\\
  \hline
   Spatial  & 0.77 & 0.56 & 0.48 & 0.41 \\
  \hline
  \end{tabular}
}
\caption{Influence similarity w.r.t.\ social contexts.}
\label{tab:infcosim}
\end{table}

\medskip
From the above analysis,
we can conclude that community impact is constrained under spatial and temporal contexts.

% ==================================================================
\section{Location Prediction}
\label{sec:locpre}
% ==================================================================

Location prediction can drive compelling applications 
including location recommendation and targeted advertising.
On the other hand, it may also threat users' privacy~\cite{STBH11}.
Following the previous analysis, 
we continue to investigate whether it is possible to use community information to effectively predict users' locations,
using machine learning techniques. 
More precisely, the question we want to answer is: 
given a user's community information, 
whether he will check in at a given place at a given time.
Note that the time here is a certain hour on a certain day (Monday to Sunday).

We first list all the features in the community-based location prediction model.
Then, we present the baseline predictors.
Experimental results are described in the~end.

% ------------------------------------------------
\subsection{Community-based location predictor}
% ------------------------------------------------

To predict whether a user will visit a certain location, 
we use one of his communities' information to establish the feature vector,
i.e., the influential community of the location (see Section~\ref{sec:commob}).

\medskip
\noindent\textbf{Community related features.} Having chosen the community, we extract its following features for prediction.
\begin{itemize}
 \item Distance between the community and the location. 
 This is the distance between the location and the community's nearest frequent movement area.
 \item Community size. Number of users in the community.
 \item Number of the community's frequent movement areas. 
 \item Community's total number of check-ins. 
 \item Community connectivity. This is the ratio between the number of edges in the community and the maximal number of possible edges.
\end{itemize}

\begin{figure}[!ht]
 \centering
 \begin{minipage}[b]{0.85\columnwidth}
   \centering
   \includegraphics[width=1\columnwidth]{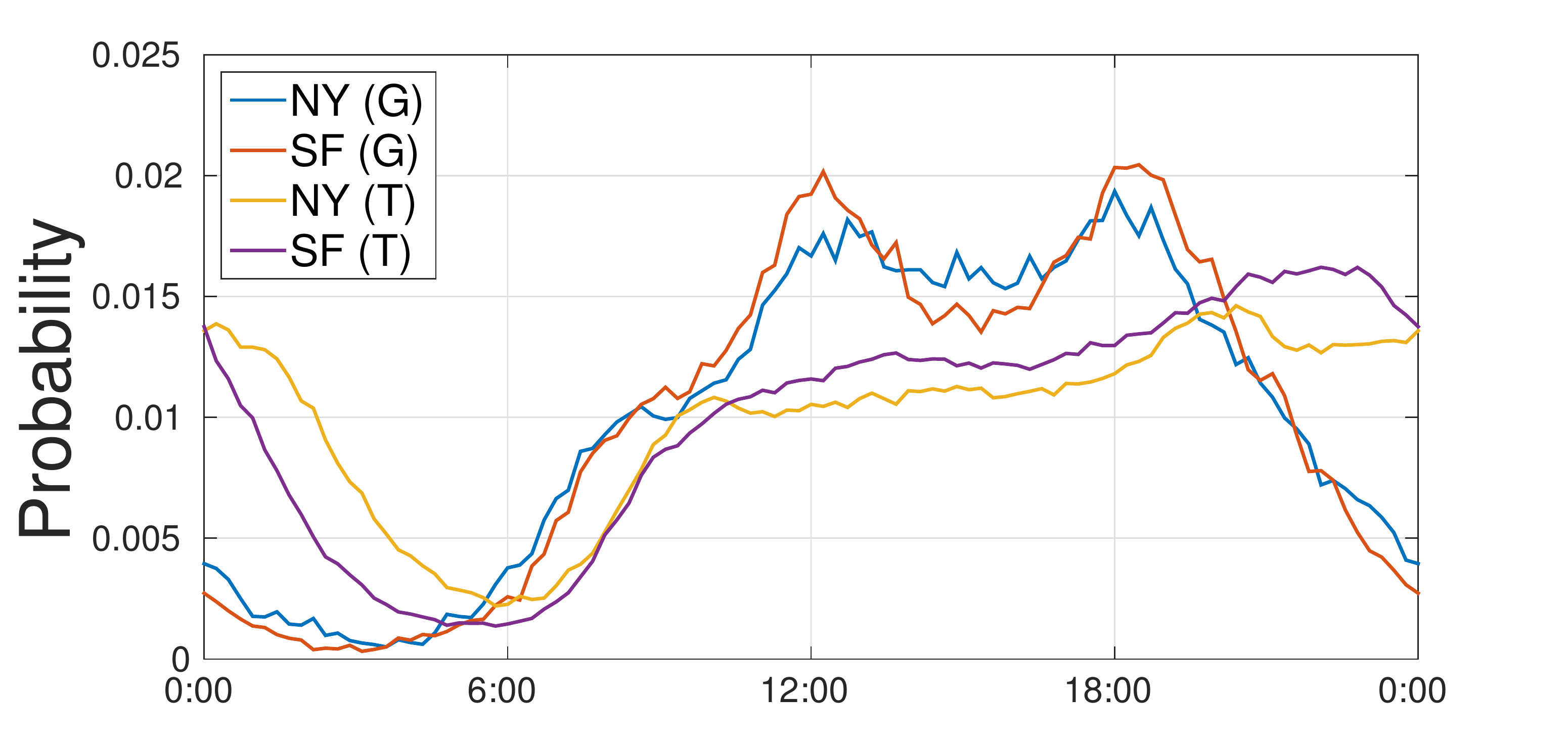}
   {\subcaption{Daily\label{fig:dtime}}}
 \end{minipage}
 
 \begin{minipage}[b]{0.85\columnwidth}
   \centering
   \includegraphics[width=1\columnwidth]{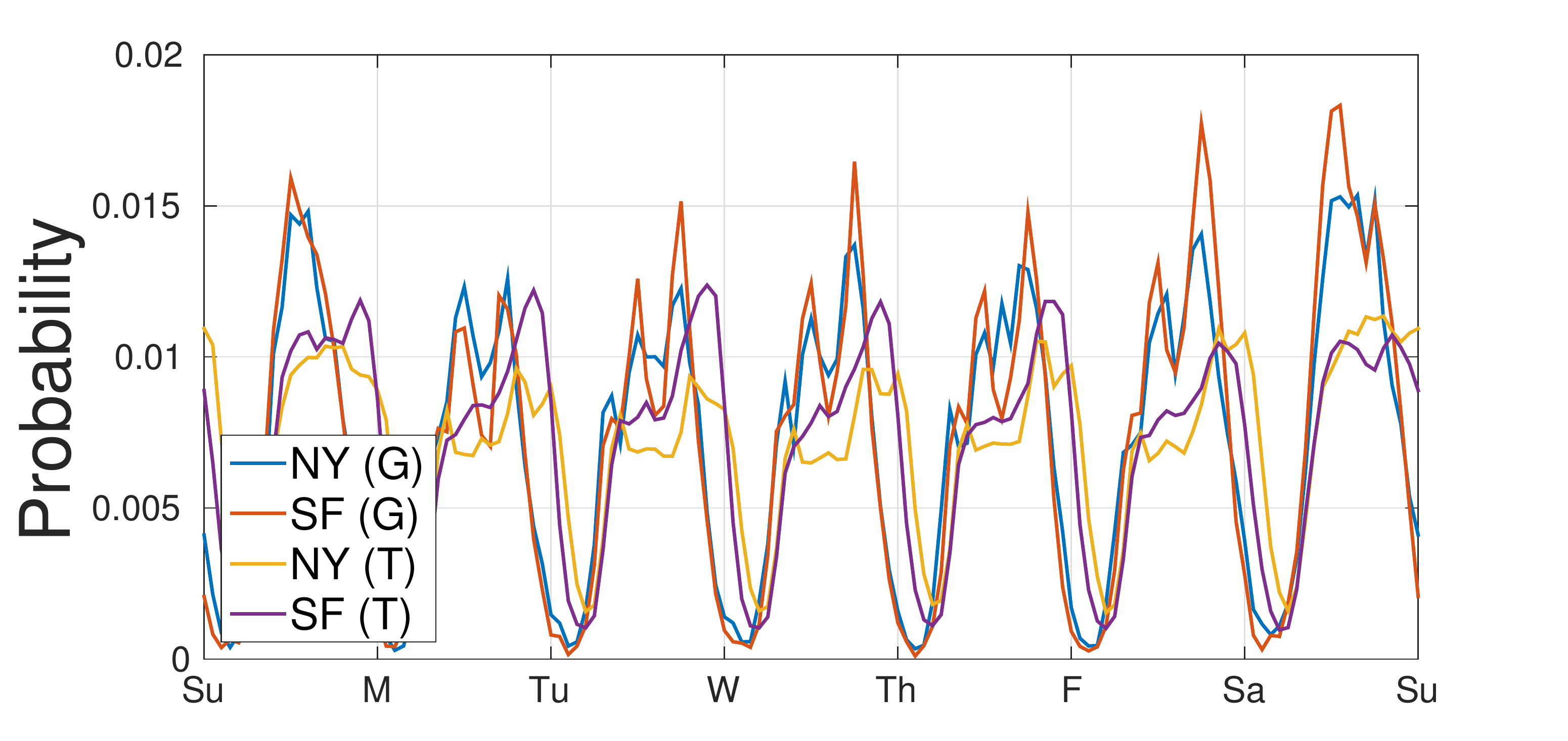}
   \subcaption{Weekly\label{fig:wtime}}
 \end{minipage}%
\caption{Check-in time in the datasets.}
\end{figure}

\noindent\textbf{Time.}
Check-ins are related to time as well.
Figure~\ref{fig:dtime} (Figure~\ref{fig:wtime}) 
plots the total number of check-ins in New York and San Francisco in a daily (weekly) scale.
Since we aim to predict whether a user will check in at a place at a certain time,
the time-related features we consider
are the total number of check-ins at the time\footnote{We consider time at a per hour unit,
thus the feature is the number of check-ins of all the users at that hour.} 
and the day (i.e., Monday to Sunday) from all users.
% Here, instead of using the exact time, we set up a two-hour range, 
% i.e., one hour earlier and one hour later than the check-in time.
% To give an example, \YZ{if we want to predict whether a user will check in at a location at 3pm on August 1st,
% then we use the number of check-ins of all users from 2pm to 4pm as one feature.
% If that day happens to be Monday, 
% then the number of check-ins of all users on Monday will be the other time-related feature.}
% % the total number of check-ins at 7:50 in the morning 
% % is the number of all check-ins performed during 6:50 and 8:50.
% \JP{Why not give the features you have used?
% For example, total number check-ins on a weekday,
% total number check-ins on 2, 4, 6 ..., 24?
% The prediction question is not for any given time, actually for
% a given hour on a weekday? Is this correct?}

 %-------------------------------------------------------------
\subsection{Baseline models}
% ------------------------------------------------

\smallskip
\noindent\textbf{Sample friends.}
In our community-based predictor, each location corresponds to the user's nearest community.
To illustrate the effectiveness of communities on predicting a user's mobility,
in the first baseline model, for each location,
we randomly sample the same number of friends as the community
and use these friends to build a ``virtual community'' (as in Section~\ref{sec:commob}).
We then replace the community related features with this virtual community's corresponding ones.
The time-related features of this model are exactly the same as the ones for the community-based model.

\medskip
\noindent\textbf{Friends.}
In the second baseline model, we consider a user's all friends instead of his communities.
The features include the shortest distance from his friends to the location and the time-related features.

\medskip
\noindent\textbf{User.}
It has been shown in~\cite{CML11,CS11} that a user's past mobility can predict his future mobility effectively.
Therefore, we also extract features from a user himself to perform prediction.
The features include the following.
\begin{itemize}
 \item The shortest distance from a user's frequent movement areas (through hierarchical clustering with cut-off distance equal to 500m) 
 to the location.\footnote{To avoid overfitting, we use half of each user's check-ins to discover his frequent movement areas 
 and the other half are used for training and testing the model.}
 \item The total number of check-ins during the day.
 \item The total number of check-ins during the hour.
\end{itemize}

\medskip
\noindent\textbf{User and community.}
In the last baseline model, we combine the features from the user's model and our community-based predictor.
 
% ------------------------------------------------
\subsection{Metrics}
% ------------------------------------------------
We partition the cities into 0.001$\times$0.001 degree latitude and longitude cells, 
a user is said to be in a cell if he has been to any place belonging to the cell.
Let {\it TP}, {\it FP}, {\it FN} and {\it TN} denote true positives, false positives, false negatives and true negatives, respectively.
The metrics we adopt for evaluation include (1) Accuracy,
\[
%\begin{array}{c}
{\it Accuracy} = \frac{\vert{\it TP}\vert + \vert{\it TN}\vert }{\vert {\it TP} \vert +\vert {\it FP} \vert + \vert {\it FN} \vert
+\vert {\it TN} \vert};
%\end{array}
\]
(2) F1 score,
\[
%\begin{array}{c}
{\it F1} = 2\cdot\frac{{\it Precision}\times {\it Recall}}{{\it Precision} + {\it Recall}},~\mbox{with}\\
%\end{array}
\]
\[
%\begin{array}{c}
{\it Precision} = \frac{\vert{\it TP}\vert }{\vert {\it TP} \vert +\vert {\it FP} \vert},~{\it Recall} = \frac{ \vert{\it TP}\vert }{\vert {\it TP} \vert +\vert {\it FN} \vert};
%\end{array}
\]
and (3) AUC (area under the ROC curve).

% ------------------------------------------------
\subsection{Experiment setup}
% ------------------------------------------------
We build a classifier for each user.
A classifier needs both positive and negative examples.
So far we only have the positive ones, i.e., a user visits a location.
To construct the negative examples,
for each location a user visits, 
we randomly sample a different location (within the city) as the place that he does not visit at that moment.
In this way, a balanced dataset for each user is naturally formed.
As in the data analysis, we only focus on active users who have at least 100 check-ins in the city.
For each user, we sort his check-ins chronologically 
and put his first 80\% check-ins for training the model and the rest 20\% for testing.
The machine learning classifier we exploit here is logistic regression.
In all sets, we perform 10-fold cross validation.

% ------------------------------------------------
\subsection{Results}
% ------------------------------------------------

\begin{figure*}[!ht]
\centering
%--------------------------------------------------------------------
  \begin{minipage}[t]{0.65\columnwidth}
    \includegraphics[width=1\columnwidth]{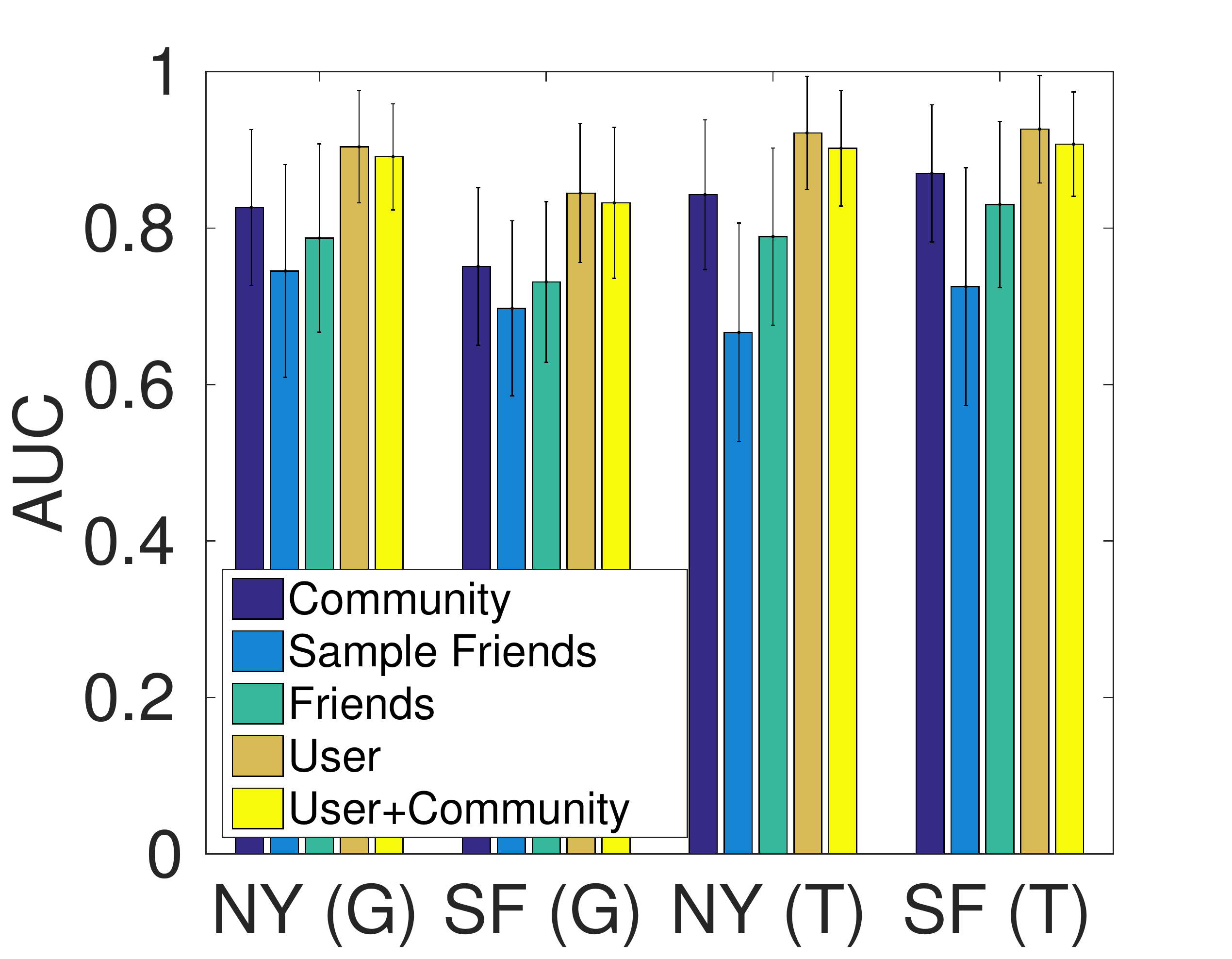}
    {\subcaption{AUC \label{fig:exp_auc}}}
  \end{minipage}
  \begin{minipage}[t]{0.65\columnwidth}
    \includegraphics[width=1\columnwidth]{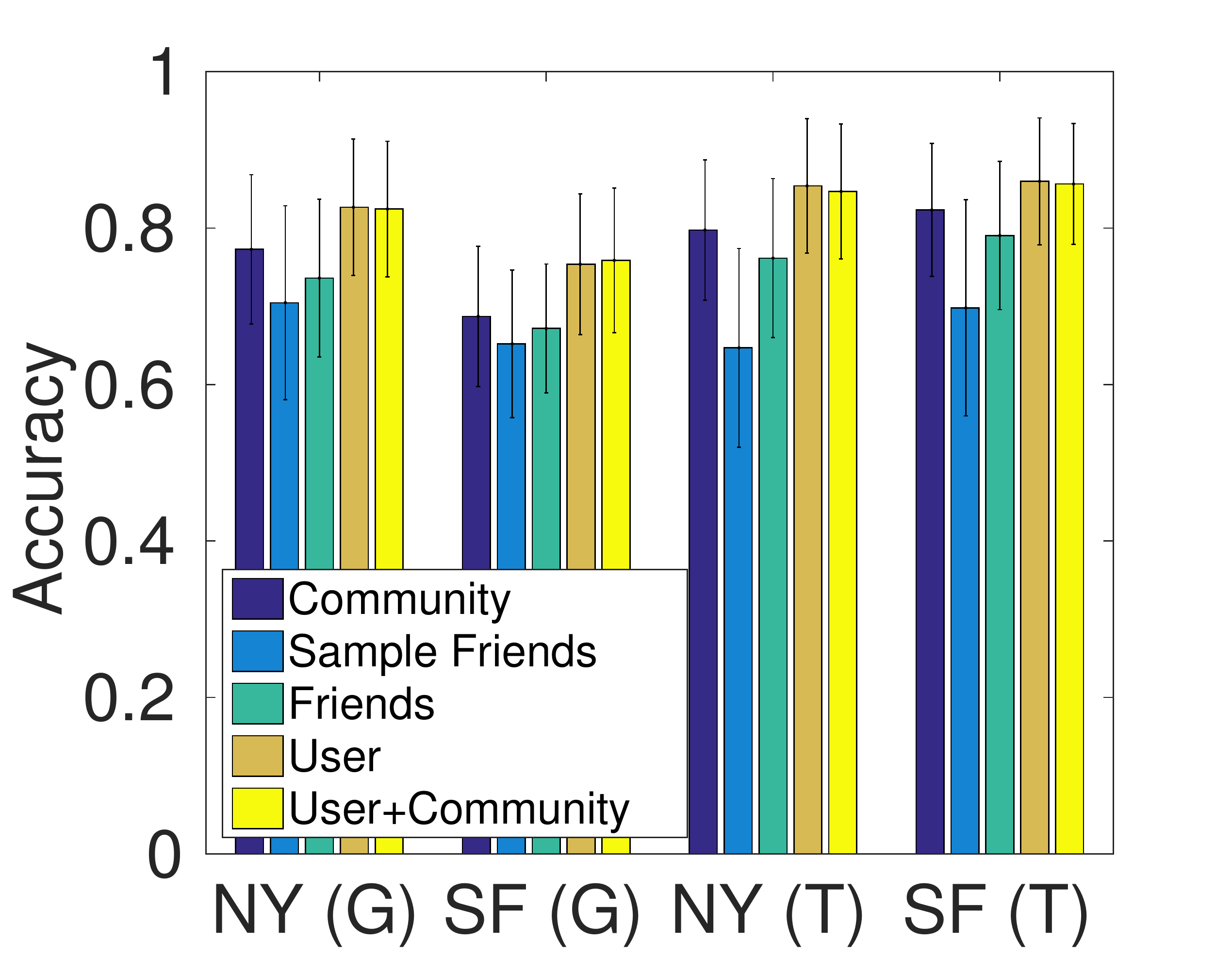}
    {\subcaption{Accuracy \label{fig:exp_acc}}}
  \end{minipage}
  \begin{minipage}[t]{0.65\columnwidth}
    \includegraphics[width=1\columnwidth]{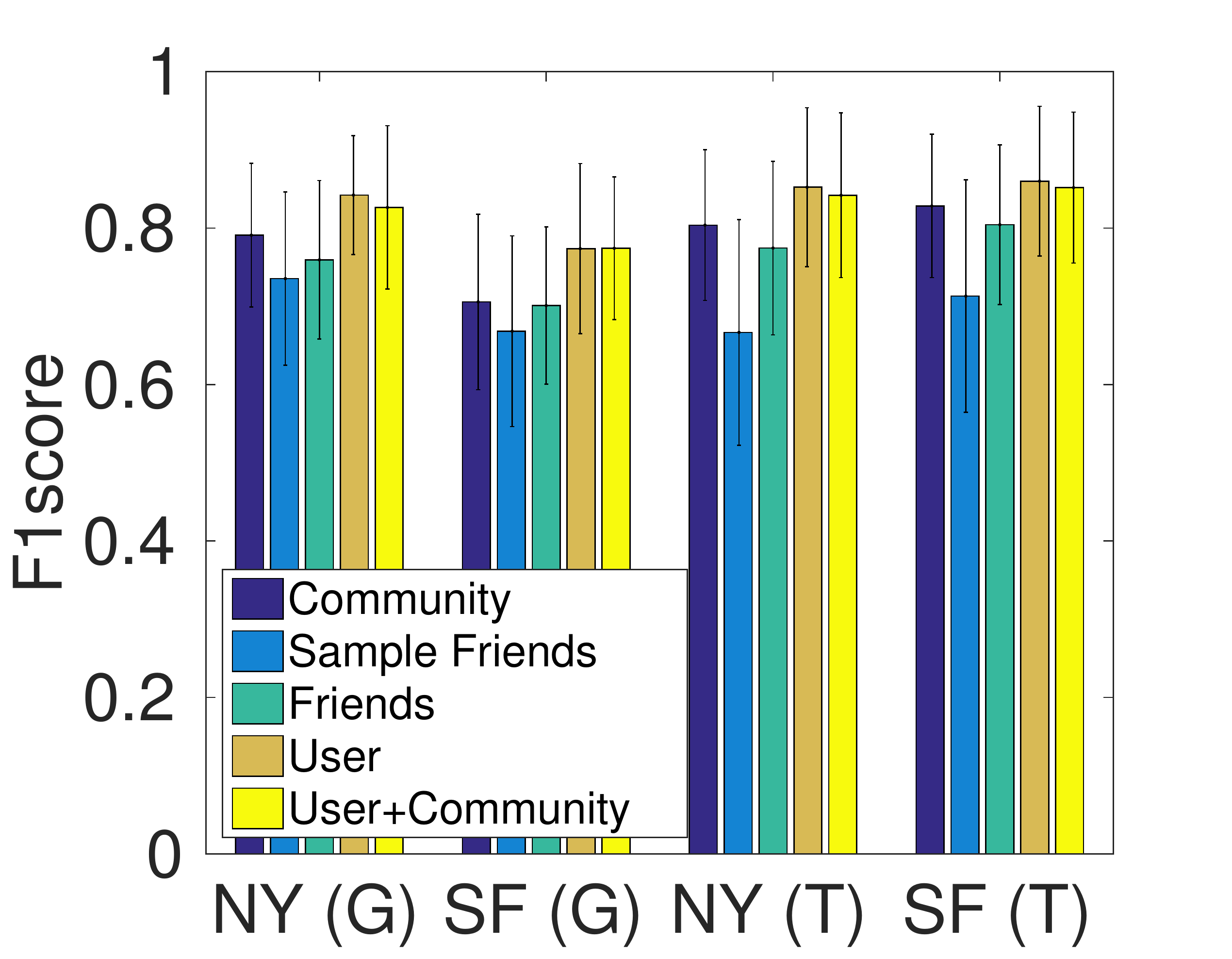}
    {\subcaption{F1score \label{fig:exp_f1score}}}
  \end{minipage}
%-------------------------------------------------------------
\caption{Prediction results.\label{fig:exp}}
\end{figure*}

% \begin{figure*}[!ht]
% \centering
% %--------------------------------------------------------------------
%   \begin{minipage}[t]{0.51\columnwidth}
%     \includegraphics[width=1.1\columnwidth]{pic/exp_aucnyg}
%     {\subcaption{NY (G) (r = 0.60) \label{fig:aucnyg}}}
%   \end{minipage}
%   \begin{minipage}[t]{0.51\columnwidth}
%     \includegraphics[width=1.1\columnwidth]{pic/exp_aucsfg}
%     {\subcaption{SF (G) (r = 0.74)\label{fig:aucsfg}}}
%   \end{minipage}
%   \begin{minipage}[t]{0.51\columnwidth}
%     \includegraphics[width=1.1\columnwidth]{pic/exp_aucnyt}
%     {\subcaption{NY (T) (r = 0.88)\label{fig:aucnyt}}}
%   \end{minipage}
%   \begin{minipage}[t]{0.51\columnwidth}
%     \includegraphics[width=1.1\columnwidth]{pic/exp_aucsft}
%     {\subcaption{SF (T) (r = 0.97)\label{fig:aucsft}}}
%   \end{minipage}
% %-------------------------------------------------------------
% \caption{AUC as a function of community entropy}
% \label{fig:auccoment}
% \end{figure*} 

\smallskip
\noindent\textbf{Performance in general.}
As depicted in Figure~\ref{fig:exp},
our community-based predictor's performance is promising
and it outperforms two baseline models that exploit friends' information.
Especially for the sample friends model, the community-based model is almost 20\% better 
among all three metrics in the Twitter dataset.
By studying logistic model's coefficients, the most important feature is the distance 
between the community and the location,
followed by the community connectivity and size.

On the other hand, two predictions that are based on user's own information
perform better than our community-based predictor.
Also, the predictor combining user and community information does not improve the performance.
This indicates that a user's past check-ins are the most useful information 
for predicting where he will be in the future which also validates the results proposed in~\cite{CML11,CS11}.

\begin{figure}[h]
\centering
%--------------------------------------------------------------------
  \includegraphics[width=0.75\columnwidth]{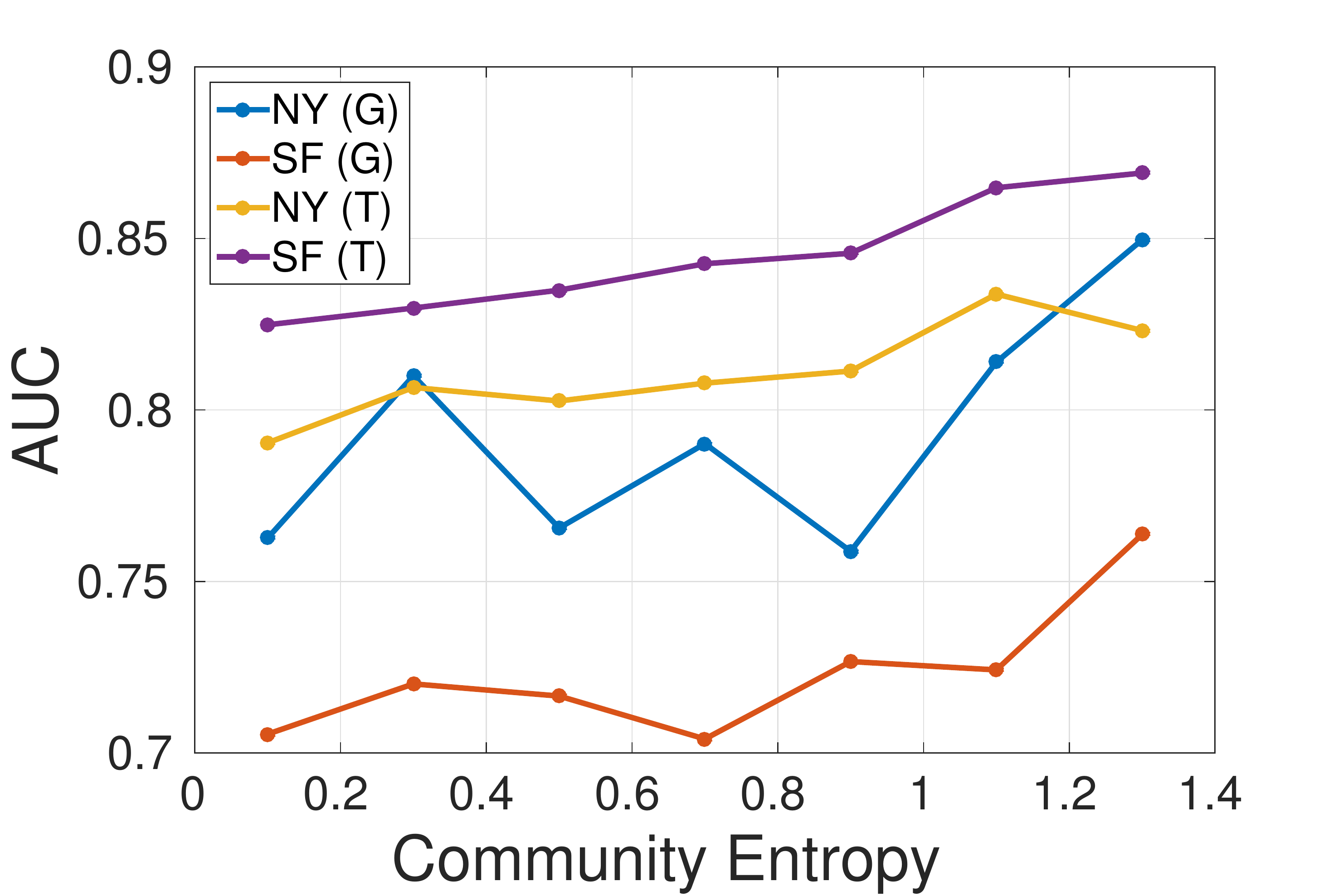}
%-------------------------------------------------------------
\caption{AUC as a function of community entropy, values of the Pearson's correlation:
0.60 (NY (G)), 0.75 (SF (G)), 0.88 (NY (T)), 0.97(SF (T)).}
\label{fig:auccoment}
\end{figure}

\medskip
\noindent\textbf{Prediction vs.\! community entropy.}
In Figure~\ref{fig:auccoment}, we bucket community entropy by intervals of 0.2
and plot its relationship with the prediction results (AUC).
As we can see, with the increase of community entropy, 
the AUC grows for the community-based model 
which means the predictor works better for users with high community entropies.
For example, the AUC value increases more than 5\% in San Francisco in the Gowalla dataset (community entropy from [0, 0.2) to [1.2, 1.4)).

We further calculate the Pearson's correlation coefficient\footnote{Pearson's correlation coefficient
is the covariance of two variables divided by the product of their standard deviations.} 
between community entropy and our prediction results.
In the Twitter dataset, the correlation coefficient for New York and San Francisco is 0.88 and 0.97 respectively,\footnote{The
two values are slightly smaller for the Gowalla dataset,
which is probably due to the fact that the Twitter dataset
contains more information on social relations than
the Gowalla dataset (see discussions in Section~\ref{sec:comdet}).}
indicating that community entropy and the prediction results are strongly correlated.
This validates our intuition that a user with high social diversity is clearly influenced by his communities.
We can conclude that 
community information can be explored to achieve promising location predictions,
especially for those users with high community entropies.

\medskip
\noindent\textbf{Difference between cities.}
In Figure~\ref{fig:exp} and Figure~\ref{fig:auccoment}, we observe
that the prediction results are different between two cities.
New York has the better performance than San Francisco in the Gowalla dataset.
On the other hand, 
the prediction results are similar in the Twitter dataset.
The reason for different performances in different cities could be due to the density of the cities 
(e.g., New York's population density is higher than San Francisco),
or the adoption of LBSN services by users in different cities.
We leave the investigation as a future work.

% ------------------------------------------------
\subsection{Other strategies to choose communities}
% ------------------------------------------------
So far, we have shown that exploring community information can lead to effective location prediction.
The community we choose is the one that has the closest frequent movement area to the target location.
% This involves collecting check-ins from all users' friends and clustering them community by community.
We would like to know if other strategies to choose community can achieve similar results.
We consider three strategies
including choosing the community with most users ({\sf max-size}), 
the community with highest connectivity ({\sf max-con}) and random community ({\sf random}).
Table~\ref{tab:comchoose} summarizes the prediction performances in New York in the Twitter dataset.
% \JP{Or Gowalla dataset? Is the table recomputed for Twitter?}\YZ{yes it is}
As we can see, our original strategy outperforms these three.
Among these three strategies, 
{\sf max-con} performs slightly better than the other two, 
but it is still relatively worse than our original strategy to choose community.
This again validates our observation in Section~\ref{sec:commob} that
influential communities are constrained by contexts (spatially or temporally),
in other words one community cannot influence every location of the user.
% Note that the results in Table~\ref{tab:comchoose} is recomputed from the Gowalla dataset as well.

\begin{table}[!ht]
\centering
\scalebox{1.0}{
\begin{tabular}{| c | c | c | c |}
  \hline
    & AUC & Accuracy & F1score\\
  \hline
  \hline
  Community & 0.83 & 0.78 & 0.79\\
  \hline
   {\sf max-size} & 0.73 & 0.72 & 0.74\\
  \hline
   {\sf max-con} & 0.74 & 0.73 & 0.74\\
  \hline
   {\sf random} & 0.71 & 0.71 & 0.72\\
  \hline
  \end{tabular}
}
\caption{Performance of community-choosing strategies.\label{tab:comchoose}}
\end{table}

% ------------------------------------------------
\subsection{Comparison with the PSMM model}
% ------------------------------------------------
In~\cite{CML11}, the authors establish a mobility model (PSMM) for each user based on his past check-ins.
The assumption behind this model is that a user's mobility is mainly centered around two states such as home and work.
Each state is modeled as a bivariate Gaussian distribution and the total mobility 
is then formalized into a dynamic Gaussian mixture model with time as an independent factor.
The check-ins that do not fit well with the two states are considered as social check-ins 
and are modeled through another friends-based distribution.
We implement the PSMM model and compare its performance with our community-based predictor.
Each user's first 80\% check-ins are used for training his PSMM model.
For testing, besides the rest 20\% check-ins, 
we also construct the same number of locations that the user does not go at the moment (as our classification setup).
As the PSMM model's output is the exact location of the user,
we consider the prediction is correct when the output location is within 1km of the real location.
Table~\ref{tab:comppmm} shows the accuracy between our model and PSMM.
In all the datasets, our community-based predictor significantly outperforms PSMM.
As suggested in~\cite{SKB12}, 
this is probably because two states are not enough to capture a user's mobility in a city.
Moreover, a user's check-in data is also too sparse to train a good PSMM model.
We leave the further investigation as a future work.

\begin{table}[!ht]
\centering
\scalebox{0.95}{
\begin{tabular}{| c || c | c || c | c |}
  \hline
   & NY (G) & LA (G) & NY (T) & SF (T)\\
  \hline
  \hline
  Community & 0.76 & 0.67 & 0.78 & 0.81\\
  \hline
  PSMM & 0.55 & 0.60 & 0.67 & 0.65\\
  \hline
  \end{tabular}
}
\caption{Comparison with PSMM on prediction accuracy.\label{tab:comppmm}}
\end{table}

% ==================================================================
\section{Related Work}
\label{sec:rel}
% ==================================================================

Thanks to the emerging of LBSNs,
mobility as well as its connection with social relations have been intensively studied~\cite{CCLS11,SNLM11,GTL12}.
There are mainly two directions of research going on in the area.
One direction is to use the location information from LBSNs to predict
friendships (see e.g.~\cite{LZXCLM08,CTHKS10,CBCSHK10,CS11,SKB12,PSL13,ZP15b}),
the other studies the impact from friendships on locations~\cite{BSM10,CS11,CML11,SKB12,MCC13}
which is what we focus on in the current work.

Backstrom, Sun and Marlow~\cite{BSM10} study the friendship and location
using the Facebook data with user-specified home addresses.
They find out that the friendship probability as a function of home distances follows a power law,
i.e., most of friends tend to live closely.
They also build a model to predict users' home location based on their friends' home.
Their model outperforms the predictor based on IP addresses.
The authors of~\cite{CS11} use the Facebook place data to study check-in behaviors and friendships.
They train a logistic model to predict users' locations.
Besides that, they also investigate how users respond to their friends' check-in 
and use the location data to predict friendships.
Cho, Myers and Leskovec~\cite{CML11} investigate the mobility patterns 
based on the location data from Gowalla, Brightkite 
as well as data from a cellphone company.
Based on their observation, they build a dynamic Gaussian mixture model for human mobility 
involving temporal, spatial and social relations features.
Sadilek, Kautz and Bigham~\cite{SKB12} propose a system for both location and friendship prediction.
For location prediction, 
they use dynamic Bayesian networks to model friends' locations (unsupervised case)
and predict a sequence of locations of users over a given period of time.
McGee, Caverlee and Cheng~\cite{MCC13} introduce the notion of social strength 
based on their observation from the geo-tagged Twitter data and 
incorporate it into the model to predict users' home locations.
Experimental results show that their model outperforms the one of~\cite{BSM10}.
Jurgens in~\cite{J13} proposes a spatial label propagation algorithm to infer a user's location 
based on a small number initial friends' locations.
Techniques such as exploiting information from multiple social network platforms
are integrated into the algorithm to further improve the prediction accuracy.

The main difference between previous works and ours is the way of treating friends.
We consider users' friends at a community level while most of them treat them the same 
(except for the paper~\cite{MCC13} which introduce `social strength',
which is based on common features but not on communities).
Moreover, our location predictor doesn't need any user's own information but his friends' 
to achieve a promising result,
especially for users' with high community entropies.
Other minor differences include the prediction target:
we want to predict users' certain locations in the future not their home~\cite{BSM10,MCC13,J13} 
or a dynamic sequences of locations~\cite{SKB12}.

We focus on understanding users' mobility behavior from social network communities.
The authors of~\cite{BNSNM12} tackle the inverse problem,
i.e., they exploit users' mobility information to detect communities.
They first attach weights to the edges in a social network based on the check-in information,
then the social network is modified by removing all edges with small weights.
In the end, a community detection algorithm (louvain method\cite{BGLL08}) is used on the modified social graph to discover communities.
The experimental results show that their method is able to discover more meaningful communities, such as place-focused communities,
compared to the standard community detection algorithm.

More recently, Brown et al.~\cite{BLMNB14} analyze mobility behaviors 
of pairs of friends and groups of friends (communities).
They focus on comparing the difference between individual mobility and group mobility.
For example, they discover that
a user is more likely to meet a friend at a place where they have not visited before;
while he will choose a familiar place when meeting a group of friends.
%
%the place type is also correlated to the mobility,
%friends tend to meet at bars not train stations.
%This work can be considered as a complement of our current work.
%In the future, we are interested in validating their results with our dataset.

% ==================================================================
\section{Conclusion and Future Work}
\label{sec:conclu}
% ==================================================================

In this paper, we have studied the community impact on user's mobility.
Analysis leads us to several important conclusions:
(1) communities have a stronger impact on users' mobility;
(2) each user is only influenced by a small number of communities;
and (3) different communities have influences on mobility under different spatial and temporal contexts.
Based on these, we use machine learning techniques to predict users' future locations 
focusing on community information.
The experimental results on two types of real-life social network datasets
are consistent with our analysis
and show that our prediction model is very effective.
The scripts for conducting the analysis and experiments as well as the Twitter dataset are available upon request.\footnote{
Preliminary results of this work are reported as a poster~\cite{PZ15}.}
% at \url{http://satoss.uni.lu/yang/communities.zip}.

In the future, we plan to extend our work in several directions.
First, we have shown in this paper that communities can be exploited to achieve a promising location prediction.
We are also interested in extending our work 
to other applications such as location recommendation.
It is possible to redesign the cost function in matrix factorization based methods for location recommendation 
by taking into account community information.
Second, we would like to conduct the analysis of community impact on other social behaviors
such as information sharing or interests adoption.
% Second, so far we only focus on the impact from community on mobility, 
% it is also interesting to pursue the opposite direction 
% that is to discover communities based on users' mobility information.
% Community detection can be treated as a clustering process,
% thus we plan to combine spatial and temporal information from check-in data
% or geo-tagged tweets together 
% to define a meaningful distance between two users.
% With this new distance, clustering algorithms such as hierarchical clustering or K-means can be applied to discover communities.
Third, in a broader point of view, our current work 
is actually a demonstration of the communities' effect on human behaviors.
As pointed by~\cite{YML14}, community is the most meaningful resolution to study social network.
Therefore, we also plan to investigate a user's role in his social network 
based on the structure of his communities.

% ==================================================================

\bibliographystyle{abbrv}
\bibliography{comloc_cosn.bib}  % sigproc.bib is the name of the Bibliography in this case

\end{document}